# Quantitative and functional post-translational modification proteomics reveals that TREPH1 plays a role in plant thigmomorphogenesis

Kai Wang [a,1], Zhu Yang [a,b,1], Dongjin Qing [a,1], Feng Ren [a,1], Shichang Liu [a], Qingsong Zheng [a,c], Jun Liu [d], Weiping Zhang [d], Chen Dai [c], Madeline Wu [a], E. Wassim Chehab [e], Janet Braam [e], and Ning Li [a,b,*]

[a] Division of Life Science, Energy Institute, Institute for the Environment, The Hong Kong University of Science and Technology, Hong Kong SAR, China.
[b] Shenzhen Research Institute, The Hong Kong University of Science and Technology, Shenzhen, Guangdong, 518057, China.
[c] Proteomics Center, College of Resources and Environmental Science
Nanjing Agriculture University, Nanjing 210095, China.
[d] ASPEC Technologies Limited, Beijing, 100101, China.
[e] Department of BioSciences, Rice University, Houston, TX, 77005, USA.
[1] These authors contributed equally to this work
* Address correspondence to Ning Li, Division of Life Science, The Hong Kong University of Science and Technology, Clear water bay, Hong Kong SAR, China. Phone: +852 2358 7335. Email, boningli@ust.hk

## Abstract

Plants can sense both intracellular and extracellular mechanical forces and can respond through morphological changes. The signaling components responsible for mechanotransduction of the touch response are largely unknown. Here, we performed a high-throughput SILIA (*stable isotope labeling in Arabidopsis*)-based quantitative phosphoproteomics analysis to profile changes in protein phosphorylation resulting from 40 seconds of force stimulation in Arabidopsis *thaliana*. Of the 24 touch-responsive phosphopeptides identified, many were derived from kinases, phosphatases, cytoskeleton proteins, membrane proteins and ion transporters. TOUCH-REGULATED PHOSPHOPROTEIN1 (TREPH1) and MAP KINASE KINASE 2 (MKK2) and/or MKK1 became rapidly phosphorylated in touch-stimulated plants. Both TREPH1 and MKK2 are required for touch-induced delayed flowering, a major component of thigmomorphogenesis. The *treph1-1* and *mkk2* mutants also exhibited defects in touch-inducible gene expression. A non-phosphorylatable site-specific isoform of TREPH1 (S625A) failed to restore touch-induced flowering delay of *treph1-1*, indicating the necessity of S625 for TREPH1 function and providing evidence consistent with the possible functional relevance of the touch-regulated TREPH1 phosphorylation. Bioinformatic analysis and biochemical subcellular fractionation of TREPH1 protein indicate that it is a soluble protein. Altogether, these findings identify new protein players in Arabidopsis thigmomorphogenesis regulation, suggesting that protein phosphorylation may play a critical role in plant force responses.

## Significance Statement

Like neural systems in animals, plants respond to a delicate force signal, such as a light touch, with extreme sensitivity, being it thigmotropism, thigmonastic movement, and thigmomorphogenesis. To understand the complex force signaling networks in plants, we applied SILIA-based quantitative PTM proteomics to measure 40 seconds of protein phosphorylation changes in Arabidopsis in response to cotton and human hair touches, and identified 4895 repeatable and non-redundant phosphopeptides, 579 of which are novel phosphosites derived from the 509 phosphoprotein groups, and finally identified 24 TOUCH-REGUALTED PHOSPHOPROTEIN (TREPHs) groups. Consequent molecular biological and bioinformatic studies revealed that both TREPH1 and MKK2 proteins are required for Arabidopsis touch response. These studies suggest that protein phosphorylation is involved in mechanotransduction of plant thigmomorphogenesis.

## Introduction

Plants perceive and respond to mechanical forces, including those resulting from extracellular stimuli such as wind and rain, as well as stimuli manifested during cell proliferation and the differential growth of neighboring cells [1-7]. Plants respond to mechanostimulation throughout growth and development [2,8], and these responses are thought to play essential roles in pattern formation [9]. Mechanical signals derived from cellular activities are integrated with the gravitational forces that are constantly imposed on plants through the weight of cellular components, such as amyloplasts and the central vacuole [10,11]. Responses to gravitational forces include tropisms, whereby many shoots grow away from the gravitational pull of the earth and roots grow towards this force [12]. Reaction wood in conifer trees and many flowering plants is produced when woody stem tissues differentially expand due to the gravitational force imposed by the weight of the stem [13]. Some plants display thigmotropic (*thigmo* means touch in Greek) growth, whereby a touch stimulus results in growth directed towards or away from the stimulus contact point; an exemplary thigmotropic response is the directed coiling of a climbing plant tendril around a supporting object [14]. Touch-induced plant movements that occur in a direction independent of the stimulus direction are called thigmonastic responses [15], and include the dramatic rapid touch-induced leaf movements of *Mimosa pudica* and the carnivorous plant Venus flytrap [16]. Thigmomorphogenesis is a slower touch response that affects overall plant growth [17]. Thigmomorphogenesis can be quite dramatic and is widespread among plants. Thigmomorphogenesis can result from diverse environmental factors, such as wind, rainfall, hail, animal contact, and even the touch of a plant organ itself [13,18]. Frequent touch stimulation of *Arabidopsis thaliana* leaves triggers delayed flowering [2,19], and long-term wind-entrained trees can become stocky [20].

Specialized mechanosensory cells or structural appendages of plants may participate in sophisticated mechanoresponses. For example, specialized endodermal and columella cells can sense changes in orientation relative to the gravity vector [21-24]. Venus flytrap modified leaves and mimosa petioles harbor trigger-hair gland cells and pulvinar cells, respectively, which perceive touch signals [2]. Root tips are also highly responsive to mechanostimulation; upon encountering growth barriers, specialized mechano-sensing founder cells initiate a barrier avoidance response that can promote lateral root initiation [25].

How plants sense various mechanical stimuli and transduce signals to regulate their diverse responses to touch signals remains elusive [2,9]. One class of potential plant mechanoreceptors have been identified based on studies of homologs of well-characterized mechanosensitive ion channels from microbes and animals [26-28]. For example, plant Mechanosensitive channel of Small Conductance (MscS)-Like (MSL) ion channels, Mating-Induced Death1 (MID1)-Complementing Activity (MCA) calcium channels, and Two-Pore potassium (K) (TPK) channels are homologs of those in bacteria, yeasts, and animals, respectively [29,30]. Recently, *Piezo*, a mechanosensitive ion channel identified in humans, was found to have plant homologs that may function as selective calcium ion channels that are responsive to a wide array of force stimuli [31,32]. Examples of these homologs include MSL8, which functions in pollen hydration and seed germination [8], and MCA1/2 proteins, which are stretch-activated and function as mechanosensitive cation channels in Arabidopsis, promoting calcium fluctuations upon mechanical loading. Another class of potential mechanoreceptors consists of multimeric protein complexes that span the plasma membrane [30]; these include the Epithelial Na+

Channel (ENaC) protein complex of the sodium channel superfamily and MEC/DEG channels [33]. A subtype of these tethered mechanosensitive channel complexes is the Transient Receptor Potential (TRP) cation channels, a superfamily that includes numerous non-voltage-gated $Ca^{2+}$ channels such as TRPN [34]. A third class of multiprotein mechanosensors includes cadherins and integrins, which span the plasma membrane and physically interconnect the intracellular actomyosin cytoskeleton with extracellular matrix (ECM) fibronectin proteins [6]. This group of multiprotein mechanosensors is usually associated with RhoA/Rho-associated coiled coil-containing kinase (ROCK), Focal Adhesion Kinase (FAK), Scr tyrosine kinase, and Rho/Ras kinases, which function as mechanosensive motors that phosphorylate downstream protein substrates upon mechanical force loading [6,35]. These proteins convert force signals into diverse downstream ion currents and biochemical signals and induce protein-protein interactions, leading to touch-inducible gene expression and mechanoresponses.

Due to the potential functional redundancy and heteromeric nature of these force receptors, forward genetic screening for components of the mechanosensory and mechanoresponse machinery may not be sufficient for identifying their functions [3,9]. To circumvent this problem, we used a high-throughput stable isotope labeling in Arabidopsis (SILIA)-based [36] quantitative post-translational modification (PTM) proteomic approach to profile touch-induced and rapidly phosphorylated sites of plant proteins to identify potential force signaling components that are important for Arabidopsis thigmomorphogenesis. We successfully identified 24 touch-responsive phosphopeptides. To initially investigate the touch response relevance of these candidates, we chose two candidate phosphoproteins, TOUCH-REGULATED PHOSPHOPROTEIN1 (TREPH1) and MITOGEN-ACTIVATED PROTEIN (MAP) KINASE KINASE2 (MKK2) to investigate further. Our initial findings suggest that both TREPH1 and MKK2 are required for a touch-induced growth response and identify a specific requirement for the amino acid S625 of TREPH1 that is phosphorylated upon plant touch stimulation. Hence, we propose an important role for protein phosphorylation dynamics in force response signaling in Arabidopsis.

## Results

### *Touch delays bolting time in Arabidopsis*

In Arabidopsis, touch induces delayed bolting and phenotypic changes in rosette leaf number, average rosette radius, and inflorescence stem height [2,19], and this response is associated with signal-specific calcium signatures [37]. To elucidate the dosage effect of touch on morphogenesis, we treated 5 groups of Arabidopsis plants with one touch per second for 10–80 s (Fig. 1A). A 10-s touch (10 consecutive touches, one touch per second) was sufficient to trigger a significant delay in bolting (Fig. 1A). The touch response increased with increasing treatment period, reaching the maximum at 40-s of touch treatment. We confirmed the cellular effect of a 40-s treatment in both *Pro35S::Aequorin* (*AEQ*) [37] and *ProCML39::LUC/Col-0* (*LUC*) transgenic plants. The 40-s touch treatment triggered a rise in fluorescence reporting rapid elevation of the cytoplasmic calcium concentration $[Ca^{2+}]_{cyt}$ in the *AEQ* plants and a luminescence rise reflecting increased expression of the recombinant gene *ProCML39::LUC* (Fig. 1B), consistent with evidence for *CML39* touch inducibility of expression [38]. The experimentally measured strength of

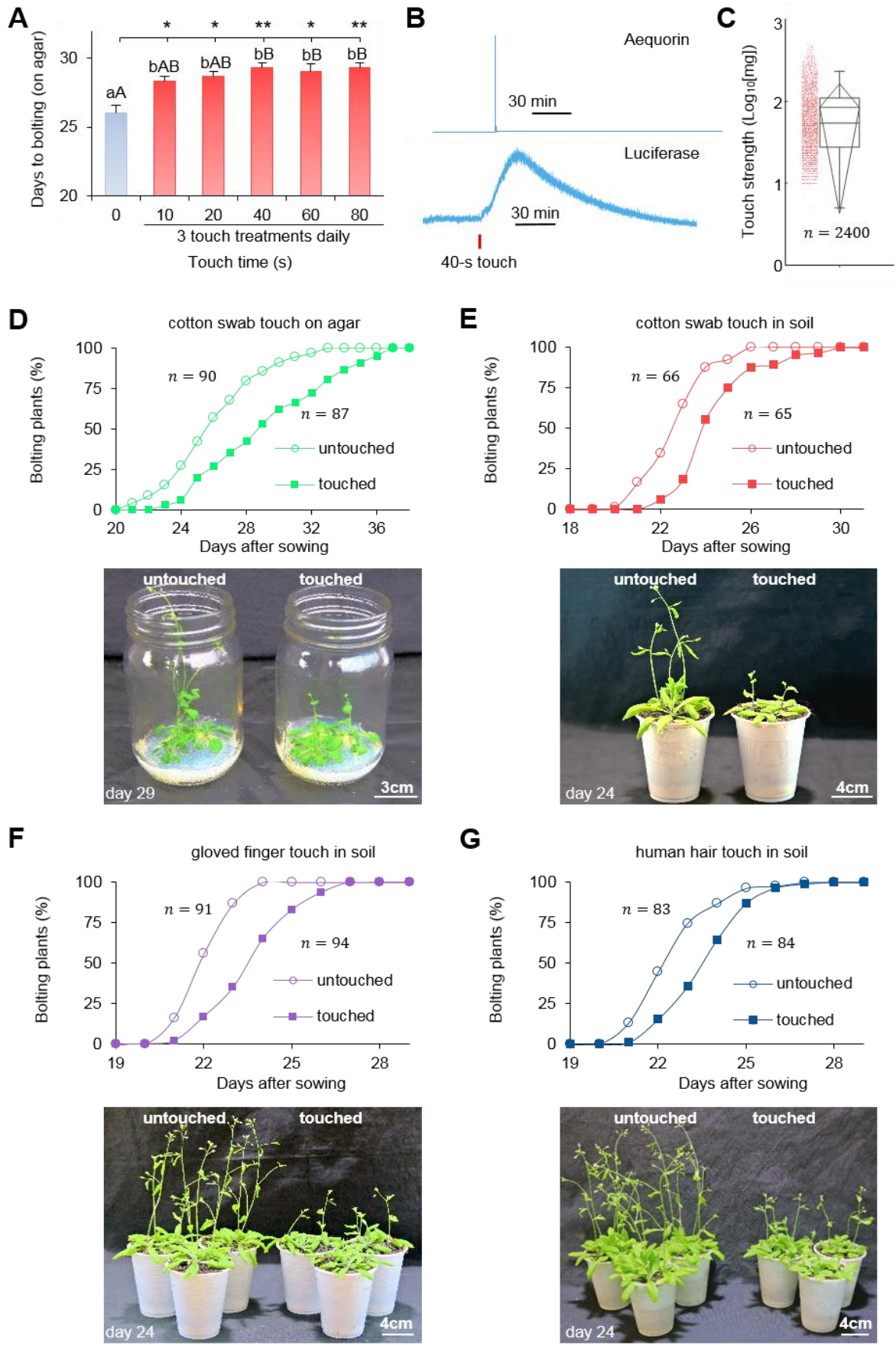

(legend on next page)

cotton swab treatment (Video S1 and S2) was 84 ± 80 mg/touch (Fig. 1C). Thus the force applied to the plants was 0.84 ± 0.8 mN/touch ($\approx$ 84 ± 80 mg/touch × 10 m/s$^2$), which was close to our targeted 1 mN (see Methods). The repetitive 40-s touch (1 touch/s) treatment with a cotton swab resulted in delayed bolting in both agar- and soil-grown plants (Fig. 1D and E). Touch treatments with similar force, repetitions, and daily frequency were also performed with soil-grown plants, using gloved finger touch (Fig. 1F) and touch administered by human-hair brushes controlled by an automated machine (Fig. 1G and Video S3); these treatments also delayed bolting.

### *Quantitative phosphoproteomic analysis of early signaling components of mechanotransduction during thigmomorphogenesis*

We analyzed the 40-s touched *LUC* plants using the SILIA-based quantitative phosphoproteomic approach (Fig. S1) [39] to identify the phosphoproteins present during early touch-response signaling. The overall workflow of this PTM proteomics is defined as 3C PTM proteomic workflow, which integrates (1) chromatographic separation and enrichment of PTM protein and peptides followed by mass spectrometry analysis with (2) computational analysis of PTM peptides using software for identification, quantitation and statistical evaluation, followed by (3) confirmation using an immunoblot analysis and/or functional analysis *via* molecular and cellular biological approaches. To achieve this objective, we firstly selected *LUC* plants for PTM proteomics analysis, as these plants report dynamic touch responses (Fig. 1B). We employed $^{14}$N- and $^{15}$N-labeled plants as

---

**Figure 1**. **The effect of touch on Arabidopsis.**

(*A*) The dose-dependent effect of touch on bolting time using cotton swabs. Twelve-day-old *LUC* transgenic plants were subjected to three rounds of touch treatment per day. In each round, touch time lasted 0, 10, 20, 40, 60, or 80 s (1 s per touch). Means ± SE are shown. Statistical analysis was performed using Student's *t*-test and Tukey's range test. Significance at *p < 0.05* and *p < 0.01* for the pairwise *t*-test is shown as * and **, respectively. The homogeneity of variance among the results was analyzed using Tukey's range test. Different lowercase and uppercase letters represent significant differences at the 5% and 0.5% level, respectively.

(*B*) Photonic reporting of calcium flux and luciferase expression in transgenic Arabidopsis after 40 s cotton swab touch. Ten-day-old *AEQ* and 14-day-old *LUC* transgenic plants were subjected to 40 s of touch treatment (1 s per touch). Fluorescence and luciferase levels peaked within 60 s and at 30 min, respectively.

(*C*) Box-and-whisker plot of touch strengths using cotton swabs. Means ± SD are shown as dashed lines. The data points were collected from *n* = 2400 touches on 30 wild type and 30 *treph1-1* plants (40 touches per plant). Means ± SD are shown as dashed lines (84 ± 80 mg), defined as the typical touch strength or "gentle touch". The touch strength median was 54 mg, upper quartile was 110 mg, and lower quartile was 28 mg. The maximum was 232 mg, while the minimum was 5 mg. The Y-axis is shown in mg (logarithmic scale).

(*D-G*) The comparison of bolting plants grown on agar and touched by cotton swab (*D*), in soil and by cotton swab (*E*), in soil and by gloved finger (*F*), and in soil and by human hairs (*G*), respectively. The upper panels are the percentage of bolting plants over the time course of treatment. The numbers of the total individual plants (*n*) from 3 biological replicates are annotated. The lower panel is the photograph of a representative of untouched and touch-treated plants.

(*A*) and (*D-G*) are from the results compiled from three independent biological replicates, with n ≥ 27 plants per replicate except (*E*) experiment n ≥ 20. Detailed touch processes were shown in Video S1–S3.

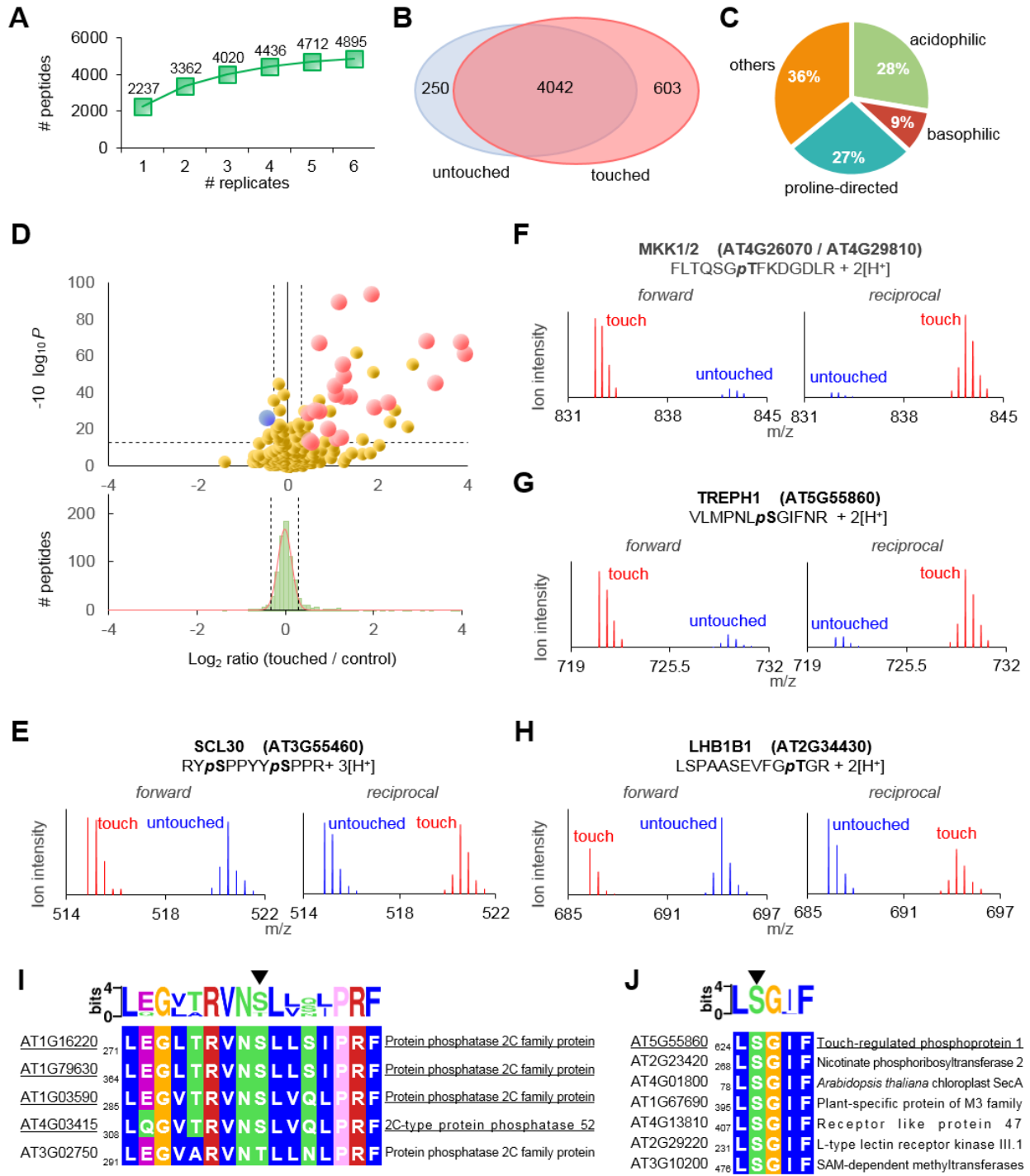

**Figure 2. Quantitative phosphoproteomic identification of touch-responsive phosphoproteins.**

(*A*) The average numbers of unique phosphopeptides observed from the combination of different sets of experimental replicates. Each replicate consisted of a mixture of $^{14}$N-labeled and $^{15}$N-labeled proteins.

(*B*) The sizes of control- and touch-specific phosphoproteomes, as well as the overlap. The mass spectrometry data are summarized in Dataset S1 and S2.

(*C*) Classification of phosphorylation sites in three general categories according to kinase docking site amino acid sequence specificity.

*(legend continued on next page)*

either the control or touch-treated plant samples in both forward and reciprocal mixing experiments (Fig. 2). A total of six biological replicates (four forward and two reciprocal replicates, see Methods) were performed with aerial tissues. As a result, this 3C PTM proteomic workflow [39] allowed us to identify 4,895 repeatable (n ≥ 2) and non-redundant (herein referred to as unique) phosphopeptides derived from 2,426 phosphoprotein groups (Dataset S1-3, false discovery rate (FDR) < 1%, Mascot Delta score ≥ 10, see Methods), which included 4616 singly, 268 doubly, 9 triply, and 2 quadruply phosphorylated peptides. Among them, 579 phosphopeptides represented novel phosphosites of the 509 protein groups (Dataset S1 and S2) according to the phosphopeptide repertoire of the PhosPhAt 4.0 database. The number of unique phosphopeptides detected from two replicates was 50% higher than that from a single replicate (3,362 vs. 2,237), whereas the number increased by only 4% (from 4,712 to 4,895) when the total number of unique phosphopeptides from six replicates was compared to that from five replicates (Fig. 2A), indicating that increasing the number of replicates to more than five contributed little to the total number of unique phosphopeptides identified using the current quantitative phosphoproteomics approach. Although over 80% of the phosphopeptides were detected in both the control and touch-treated samples, the number of touch-specific phosphopeptides (603) was much higher than that of the untouched control (250), approximately 2.4-fold (Fig. 2B). According to the substrate sequence specificity of Ser/Thr protein kinases [40], 28%, 9%, and 27% of the phosphorylated Ser/Thr are found in acidophilic ($_p$[S/T][D/E] or $_p$[S/T]xx[D/E]), basophilic (Rxx$_p$[S/T]), and proline-directed ($_p$[S/T]P) motifs, respectively (Fig. 2C). Moreover, we

---

(*D*) Volcano plot of quantitative phosphopeptide analysis. The circles in the upper panel represent quantified phosphopeptides. The Log2 ratio is the average binary logarithmic ratio of MS1 isotopologue areas of a phosphopeptide between control and touch-treated plants (Dataset S3), and *p* is the *p*-value determined using Student's *t*-test followed by Benjamini-Hochberg multiple hypothesis test correction. The lower panel shows a histogram of the Log2 ratios, which was fitted using a normal distribution (red curve). In both panels, the vertical black dashed lines indicate the mean ± 2 SD (standard deviation) of the distribution of the log2 ion intensity ratio of all phosphopeptides. The red and blue circles represent significantly touch-enhanced and -suppressed phosphopeptides, respectively, defined by *t*-test followed by Benjamini-Hochberg correction. The horizontal dashed line indicates the cutoff threshold for the significance (FDR of the BH correction, or named BH-FDR, ≤ 5%).

(*E-H*) Mass spectrograms (MS1) of sample phosphopeptides that were independently (*E*), up- (*F*), and (*G*) and downregulated (*H*) by 40-s touch, respectively. *Forward* and *reciprocal* indicate the mixing of tissues from [14]N-labeled touch-treated plants with [15]N-labeled control plant tissues and the mixing of [15]N-labeled tissues from touch-treated plants with [14]N-labeled control tissues, respectively. m/z represents the ratio of ion mass over the charge of each phosphopeptide ion. Subscript *p* marks the amino acid of a phosphosite.

(*I*) Highly conserved Touch-Regulated Phosphosite (TREPH) motifs. VN*p*SLLSIPR and VN*p*SLVQLPR, the touch-enhanced phosphopeptides of PP2C family members, were used to construct a WebLogo of a touch-regulated motif from the highly conserved amino acid sequences surrounding the phosphosites and those identified from amino acid sequence BLAST searching. See Fig. S2 for detail.

(*J*) BLAST-based phosphosite motif predicted from phosphosite S625 of TREPH1 protein.

(*I and J*) TAIR accessions and annotations are labeled on both sides of the peptides. The subscript numbers before the peptide sequences indicate the positions of the first amino acid residues in the corresponding proteins. Triangles (▼) indicate the phosphorylation sites. The experimentally identified phosphoproteins are underlined. Detailed *SILIA* framework was shown in Fig. S1.

**Table 1.** Phosphopeptides regulated by 40-s touch treatment in *Arabidopsis*

| Accession No. [a] | Fold change [b] | *p*-value [c] | Phosphopeptide [d] | Motif category | Protein annotation [e] | Subcellular localization [f] |
|---|---|---|---|---|---|---|
| **Touch-enhanced** | | | | | | |
| AT4G26070, AT4G29810 | 15.2 | 6.8×10^-09 | 23, 25 FLTQSG***p*T**FKDGDLR | [pT]xxD/E | MAP kinase kinase 1/2 (MKK1/2) | cytosol; cytosol |
| AT5G20490 | 14.2 | 8.9×10^-10 | 1240 GSPQSAGL***p*S**FLNR | | Myosin XI K (XIK) | Golgi |
| AT5G62810 | 9.7 | 4.7×10^-07 | 427 ***p*S**WVPPQPPPVAMAEAVEAIR | Rxx[S] | Peroxin 14 (PEX14) | peroxisome |
| AT5G55860 | 8.3 | 5.9×10^-10 | 619 VLMPNL***p*S**GIFNR | | TREPH1[g], Plant protein of unknown function (DUF827) | nucleus, cytosol [h] |
| AT5G10470 | 4.6 | 8.9×10^-06 | 44 RN***p*S**ISTPSLPPK | Rxx[S] | Kinesin like protein for actin based chloroplast movement 1 (KAC1) | cytosol |
| AT4G26130, AT5G59960 | 3.7 | 1.7×10^-05 | 273, 209 LD***p*S**FLR | Rxx[S] | TREPH2/3 [g], unknown protein | plasma membrane; endoplasmic reticulum |
| AT5G04870 | 3.6 | 5.3×10^-13 | 605 SF***p*S**IALKL | | Calcium dependent protein kinase 1 (CPK1) | cytosol |
| AT3G13300 | 2.6 | 3.8×10^-06 | 821 VFCSQV***p*S**NLSTEMAR | | VARICOSE (VCS) | cytosol |
| AT1G18740, AT1G74450, AT1G43630 | 2.4 | 3.3×10^-06 | 203, 211, 225 SL***p*S**WSVSR | Rxx[S] | TREPH4/5/6 [g], Protein of unknown function (DUF793) | nucleus; nucleus; nucleus |
| AT3G08510 | 2.3 | 1.8×10^-07 | 277 EVP***p*S**FIQR | | Phospholipase C 2 (PLC2) | plasma membrane |
| AT1G72410 | 2.3 | 3.0×10^-08 | 647 NL***p*S**DLSLTDDSK | Rxx[S], [pS]D/E | COP1-interacting protein-related | nucleus |
| AT1G05805 | 2.2 | 1.7×10^-03 | 186 SQL***p*S**FTNHDSLAR | | bHLH DNA-binding, ABA-responsive kinase substrate 2 (AKS2) | nucleus |
| AT3G05090 | 2.2 | 2.7×10^-12 | 377 GGSFLAGNL***p*S**FNR | | Lateral root stimulator 1 (LRS1) | plasma membrane |
| AT1G64740, AT1G50010, AT1G04820, AT4G14960 | 2.1 | 2.0×10^-03 | 340, 340, 340, 340 TIQFVDWCP***p*T**GFK | | Tubulin alpha-1/2/4/6 (TUA1/2/4/6) | cytosol; cytosol; cytosol; cytosol |
| AT3G13910 | 2.1 | 2.2×10^-06 | 58 SW***p*S**FSDPESR | Rxx[S], [pS]xxD/E | TREPH7 [g], Protein of unknown function (DUF3511) | cytosol |
| AT4G35310, AT2G17290 | 2.0 | 7.8×10^-07 | 547, 535 NSLNI***p*S**MR | [pS]xxD/E | Calmodulin-domain protein kinase 5/6 (CDPK5/6) | nucleus, cytosol; cytosol |
| AT1G24300 | 1.8 | 4.6×10^-04 | 1259 NN***p*S**LLSGIIDGGR | Rxx[S] | TREPH8 [g], GYF domain-containing protein | nucleus |
| AT5G22030 | 1.6 | 1.2×10^-09 | 303 SN***p*S**LSFLGK | Rxx[S] | Ubiquitin-specific protease 8 (UBP8) | nucleus |
| AT4G24275 | 1.6 | 3.7×10^-05 | 44 ***p*S**VSASAQAVPSPIK | Rxx[S] | Identified as a screen for stress-responsive genes | nucleus |
| AT1G16220, AT1G79630 | 1.5 | 2.5×10^-05 | 468, 481 VN***p*S**LLSIPR | Rxx[S] | Protein phosphatase 2C family protein | nucleus; nucleus |
| AT5G09890 | 1.4 | 4.3×10^-03 | 459 DTNFIGF***p*T**FK | | Protein kinase family protein | cytosol |
| AT5G20650 | 1.4 | 3.0×10^-03 | 68 SS***p*S**GVSAPLIPK | Rxx[S] | Copper transporter 5 (COPT5) | vacuole |
| AT1G03590; AT4G03415 | 1.4 | 4.9×10^-05 | 446, 452 VN***p*S**LVQLPR | Rxx[S] | Protein phosphatase 2C family protein; 2C-type protein phosphatase 52 (PP2C52) | plasma membrane; nucleus |
| **Touch-suppressed** | | | | | | |
| AT2G34430 | -1.4 | 9.4×10^-05 | 20 LSPAASEVFG***p*T**GR | | Light-harvesting chlorophyll-protein complex II subunit B1 (LHB1B1) | plastid |

[a] The multiple accession numbers sharing same phosphopeptides are homologous proteins.
[b] Fold changes of ± 1.3 (i.e., log2 change of ± 0.4) were employed as the cutoff in selecting the altered phosphopeptides.
[c] Given by two-tailed Student's *t*-test. All listed phosphopeptides were selected using two-tailed Student's *t*-test followed by Benjamini-Hochberg correction (BH-FDR < 5%).
[d] The phosphorylated amino acids (Ser or Thr) are marked by subscript *p*.
[e] The annotation is based on the description of TAIR10 (http://www.Arabidopsis.org/).
[f] The information is from the Bayes Consensus Classifier output (SUBAcon) of SUBA4 (http://suba.live).
[g] Touch-regulated phosphoproteins.
[h] The results of the present study.

identified six touch-specific phosphorylation motifs *via* bioinformatics analysis of the phosphoproteome (Fig. S2), suggesting that some phosphorylation events may be catalyzed by related kinases.

We also performed quantitative PTM proteomic analysis of the ratios of isotope-coded peptides in 750 strictly selected, quantifiable pairs of light and heavy phosphopeptides (See Methods and Dataset S3 for selection criteria), leading to the identification of 23 touch-enhanced phosphopeptides and one touch-suppressed phosphopeptide (Table 1 and Fig. 2 D-H), corresponding to a total of 24 phosphoprotein groups (in which several phosphoproteins may share the same phosphopeptide). The largest alterations (either increase or decrease) were obtained on the phosphopeptides, $_{23,25}$FLTQSG*p*TFKDGDLR (15.2-fold), derived from MAP KINASE KINASE1 (MKK1, AT4G26070) and/or MKK2 (AT4G29810), and $_{20}$LSPAASEVFG*p*TGR (-1.4-fold) from the Light-harvesting chlorophyll-protein complex II subunit B1 (LHB1B1, AT2G34430). These SILIA-based quantitative PTM proteomic results are consistent with the observed difference in the number of phosphopeptides obtained from the differential PTM proteomic analysis of both control and the touch-treated samples (*i.e.,* touched > untouched phosphopeptides in phosphopeptide number, Fig. 2B). It is therefore that touch treatment increased both the concentration and diversity of some phosphopeptides. This list of phosphopeptides (Table 1) includes MKK1 and/or MKK2, CALCIUM-DEPENDENT PROTEIN KINASE1 (CPK1, AT5G04870), CALMODULIN-DOMAIN PROTEIN KINASE5/6 (CPK5/6, AT4G35310/AT2G17290), and PROTEIN PHOSPHATASE 2C (PP2C) family proteins, which are known protein phosphorylation/dephosphorylation enzymes that function in cell signaling [41]. In addition, we identified eight touch-enhanced phosphopeptides from eight phosphoproteins of unknown function. We named these proteins <u>T</u>ouch-<u>R</u>egulated <u>P</u>hosphoproteins (TREPHs, Table 1). The levels of two similar peptides, VN*p*SLLSIPR and VN*p*SLVQLPR (derived from distinct members of the PP2C family), increased 1.8- and 1.4-fold, respectively, in response to touch treatment. These results, in combination with the results of the bioinformatics analysis, revealed a conserved touch-regulated phosphosite motif in PP2C family proteins (Fig. 2I). Bioinformatic analysis of the touch-regulated phosphosites (listed in Table 1) predicted a number of putative touch-regulated phosphosites, one of these motifs is the S625 phosphosite motif of TREPH1, which shares sequence similarity with putative phosphosites in RECEPTOR LIKE PROTEIN 47, (RLP47, AT4G13810) and L-TYPE LECTIN RECEPTOR KINASE III.1 (LECRK-III.1, AT2G29220) (Fig. 2J). These results suggest that these putative kinases might play a role in protein phosphorylation changes following plant touch stimulation.

## *TREPH1 and MKK2 phosphorylation in response to various types of force loadings, water sprinkling, and air blowing*

Following the two major steps of 3C quantitative PTM proteomics, we sought to verify the identified phosphosites using immunoblot assays (Fig. 3 and Fig. S3 and S4). Among the phosphosites with larger fold changes in phosphorylation levels (Table 1), two phosphosites attracted our interest because of the profound magnitude of phosphorylation changes and the putative functions of related phosphoprotein groups (Table 1). The first one was TREPH1, whose phosphorylation increased 8.3-fold in response to touch and its cellular and biological function was unknown, while the second one the T31 residue of MKK2 (and/or T29 of MKK1) with the greatest (15.2-fold) phosphorylation increase (Table 1). Polyclonal antibodies were generated against both the S625 non-phosphorylated (anti-

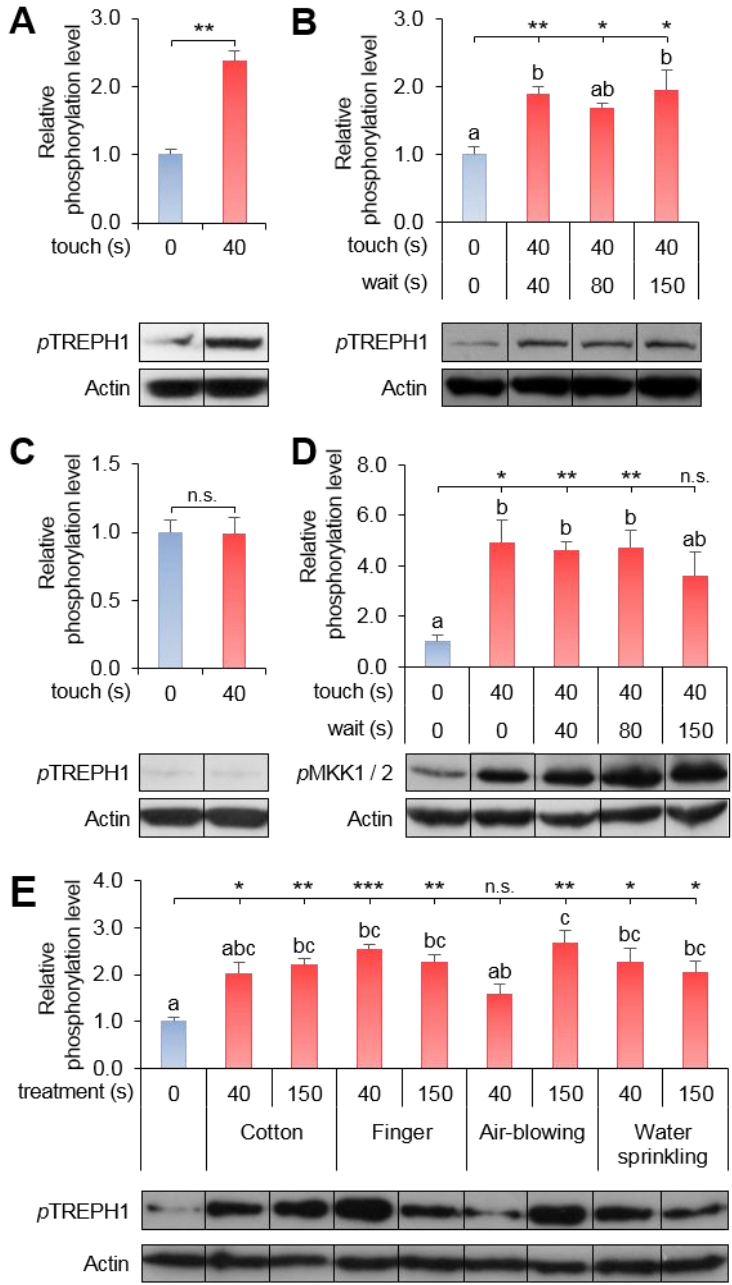

**Figure 3. Protein immunoblot analysis of the touch-regulated phosphorylation of MKK1/2 and TREPH1.**

(*A*) The enhanced phosphorylation level of TREPH1 after 40-s cotton swab touch treatment.

(*B*) Time-dependent changes in the phosphorylation of S625 of TREPH1.

(*C*) Negative control of TREPH1 S625 phosphorylation antibody in touch-treated *treph1-1*.

(*D*) Time-dependent changes in the phosphorylation of T31 of MKK2 (and/or T29 of MKK1) after 40-s cotton swab touch treatment.

All plants used in (*A-D*) were grown for 3 weeks on M/S agar medium in glass jars and performed cotton swab touch before harvested.

(*E*) Changes in phosphorylation levels at S625 of TREPH1 in Arabidopsis plants touched with a cotton swab and finger, treated with air-blowing**,** and treated with water sprinkling, respectively. The plants were grown for 3 weeks in soil.

The relative phosphorylation level was determined according to the relative level of phosphoprotein against the level of actin protein, as measured by immunoblotting using both polyclonal anti-phosphosites antibodies (anti-*p*MKK1/2 and anti-*p*TREPH1) and a monoclonal anti-actin antibody. Each bar represents the results of three immunoblotting analyses (three biological replicates). Means ± SE are shown. Statistical analysis was performed by Student's *t*-test and Tukey's range test. Significance at $p < 0.05$, $p < 0.01$, $p < 0.001$, and $p \geq 0.05$ for pairwise *t*-test are shown as *, **, ***, and n.s., respectively. The homogeneity of variance in each panel with multiple (more than two) values was analyzed using Tukey's range test. Different letters represent significant differences at the 5% level. The non-adjacent lanes from the same and whole immunoblot membranes that are shown in Fig. S3, while the multiple biological replicates for each immunoblot data point are shown in Fig. S4. The genotypes of mutants were validated and are shown in Fig. S5.

npTREPH1) and S625-phosphorylated TREPH1 (anti-pTREPH1) as well as against the T29-phosphorylated MKK1 and/or T31-phosphorylated MKK2 (anti-pMKK1/2) peptide. The antibodies were validated using immuno-dot blots with synthetic peptides (Fig. S3 A and H). Immunoblot-based quantitation of phosphorylation demonstrated that the phosphorylation of TREPH1 increased 2.4 ± 0.2 fold in response to touch (Fig. 3A and Fig. S3B and S4A), and it was sustained after a 40-s cotton-swab touch (Fig. 3B and Fig. S3C and S4B). In contrast, the touch-induced phosphorylation of TREPH1 was either undetectable or fairly low in the untouched wild-type plants and the T-DNA insertional mutant *treph1-1* (Fig. 3C and Fig. S3D and S4C; see Methods and Fig. S5 for details about genotyping). The phosphorylation level of MKK1/2 also increased (4.9 ± 0.9 fold) and was maintained after a 40-s cotton-swab touch (Fig. 3D and Fig. S3E and S4D). To verify that the touch stimulus was the causal inducer, we also touched plants with gloved fingers and analyzed the responses. A 40- or 150-s gloved-finger touch of wild-type plants led to 2.9 ± 0.3- and 2.3 ± 0.2-fold increases in pTREPH1, respectively (Fig. 3E and Fig. S3F and S4E). Similarly, TREPH1 phosphorylation increased in response to 150-s air blowing (mimicking wind) by 3.1 ± 0.3 fold (Fig. 3E and Fig. S3F and S4E), and to 40- and 150-s water treatments (resembling rainfall) by 2.3 ± 0.3 and 2.0 ± 0.2 fold, respectively (Fig. 3E and Fig. S3F and S4E). These results in indicate that diverse mechano stimuli can increase TREPH1 phosphorylation and thus confirm the 3C quantitative PTM proteomics results on TREPH1.

### TREPH1 and MKK2 are required for the touch-caused bolting delay

To examine whether the phosphorylation changes of TREPH1 and MKK2 induced by mechanical force may have functional relevance in plant touch responses, we screened homozygous T-DNA insertional mutant lines of six putative phosphoprotein-encoding genes, including *TREPH1* and *MKK2*, to determine if these mutants displayed defective bolting time responses to touch stimulation. *Mkk2* mutant was chosen for the first round of tentative genetic analysis, instead of *mkk1*, because of the availability of the mutant seed. We verified the lack of full-length transcript production from *treph1-1* together with sequencing of the mutated allele (Fig. S5). *treph1-1* also fails to show an increase in an immunoreactivity to the anti-pTREPH1 antibodies in plants treated with touch (Fig. 3C), providing additional evidence that *treph1-1* may fail to produce TREPH1 protein.

Both *treph1-1* and *mkk2* show a conspicuous defect in touch-induced delay of bolting in in comparison to the *Col-0* wild-type plants when subjected to either cotton-swab or automated human hair touch treatments (Fig. 4 A-F). *Col-0* plants exhibited 1.9 days of delay in bolting time (Fig. 4A) in response to cotton swab touch treatment, whereas delayed bolting was not observed in *mkk2* (23.2 ± 0.2 days vs. 23.4 ± 0.2 days) or *treph1-1* (23.3 ± 0.2 days vs. 23.4 ± 0.1 days) (Fig. 4 C and E and Fig. S6) under these conditions. Similarly, *Col-0* exhibited 1.5 days of delay in bolting time (Fig. 4B) in response to automated touch treatment, whereas the delayed bolting was not observed in *treph1-1* (23.9 ± 0.2 days vs. 23.9 ± 0.2 days) or *mkk2* (23.8 ± 0.2 days vs. 24.0 ± 0.1 days) (Fig. 4 D and F and Fig. S6). Two other phenotypic changes associated with the Arabidopsis touch response and thigmomorphogenesis are changes in inflorescence stem height and average rosette radius [2]. These features of thigmomorphogenesis were not abolished in *treph1-1* and *mkk2* although the touch effects on these phenotypes were reduced by the mechano-stimulus (Fig. S7). Overall, these results demonstrate that both *treph1-1* and *mkk2* are defective in touch-induced delay of bolting, and therefore that both *TREPH1* and *MKK2* are essential for this aspect of thigmomorphogenesis.

### The amino acid residue modified with phosphorylation in TREPH1 is required for touch-induced delay of flowering

Our data indicate that TREPH1 is essential for Arabidopsis to respond to force stimulation by delayed bolting. We next sought to investigate the potential role of S625

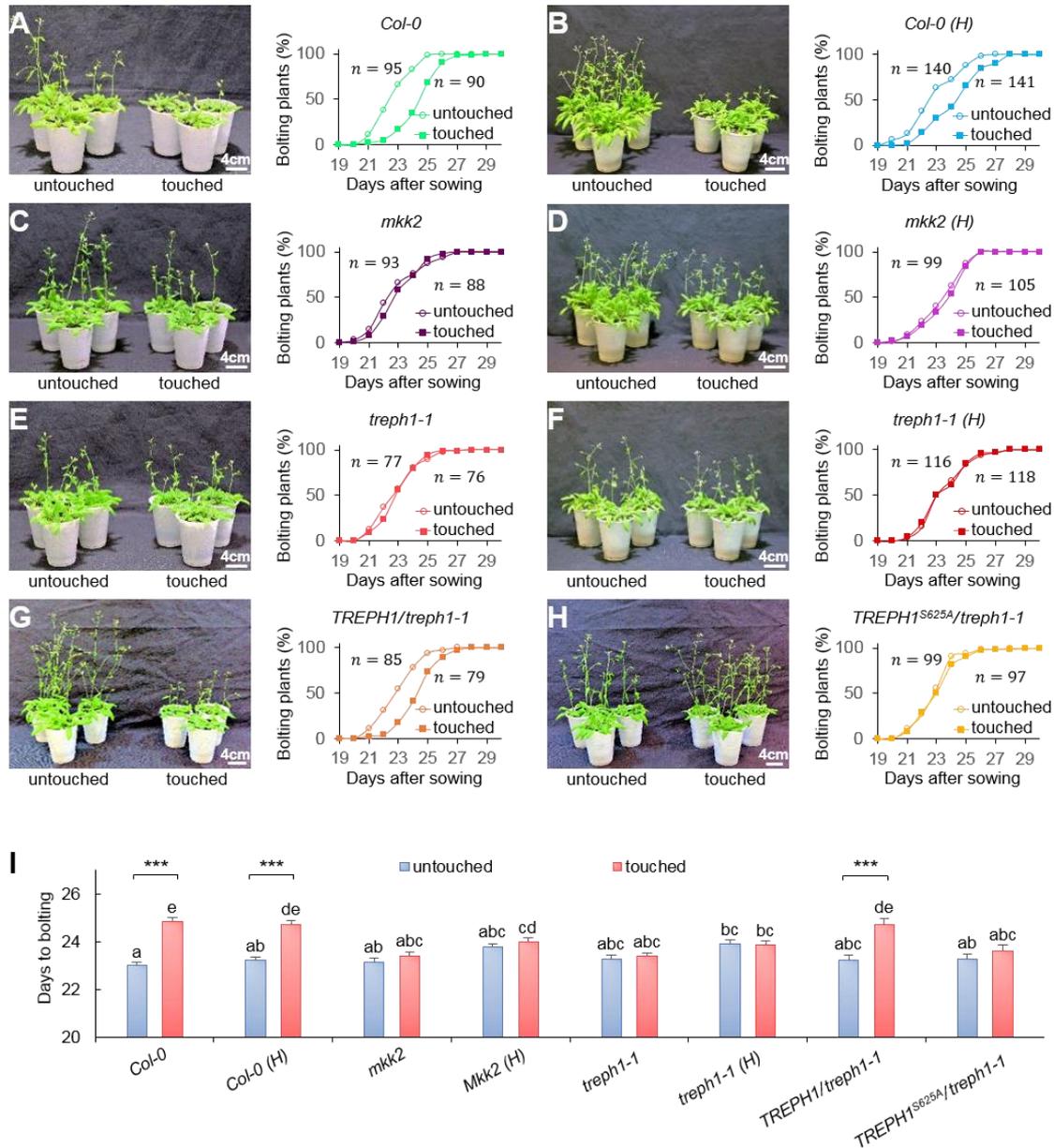

**Figure 4. Touch responses of wild-type, mutant, and transgenic plants expressing the TREPH1 isoforms.**

(*A-H*) Representatives individuals (the left panel) and the percentage of bolting individuals over time (the right panel) for the untouched control and *Col-0* cotton swab touched (*A*), *Col-0* hair touched (*B*), *mkk2* cotton swab touched (*C*), *mkk2* hair touched (*D*), *treph1-1* cotton swab touched (*E*), *treph1-1* hair touched (*F*), *TREPH1/treph1-1* cotton swab touched (*G*), and *TREPH1^S625A^/treph1-1* cotton swab touched (*H*) plants. Touch treatment began using 12-day-old plants. Each plant was subjected to three rounds of touch treatment per day (40 touches per round). H stands for human hair touch controlled by a fully automated machine. All the photos were taken between day 24 - 25 to show the difference in bolting time.

The numbers of the total individual plants (*n*) from 3 biological replicates are annotated. The numbers of individuals in each replicate are shown in Fig. S6. Additional morphological measurements were shown in Fig. S7. The genotype validation of these lines is shown in Fig. S5 and S8.

(*I*) Day to bolting in untouched and touch-treated wild-type and mutant plants. Means ± SE are shown. All results were compiled from three biological replicates (n ≥ 24, n = number of individuals per replicate). Statistical analysis was performed by Student's *t*-test and Tukey's range test. Significance between the untouched and touch-treated group of each genotype was obtained by *t*-test; *** represents *p < 0.001*. The homogeneity of variance among all samples was analyzed using Tukey's range test. Different letters represent significant differences at the 5% level.

phosphorylation of TREPH1 in this touch response. We, therefore, compared the performance of TREPH1 with and without the phosphorylatable S625 amino acid. We generated two transgenes for complementation tests of *treph1-1* (Fig. S8): *dPro35S::His::YFP::TREPH1*, which retains the capability of being phosphorylated (Fig. S3G), and *dPro35S::His::YFP::TREPH1^S625A*, which should not be phosphorylatable. These transgenes were introduced into *treph1-1* for analysis. In the presence of *dPro35S::His::YFP::TREPH1,* the touch-induced bolting time delay of *treph1-1* was rescued; touched plants bolted with an average of 24.7 ± 0.2 days compared to untouched plants, which bolted at 23.3 ± 0.2 days (Fig. 4G and Fig. S6), whereas the *dPro35S::His::YFP::TREPH1^S625A*-complemented population showed no significant delay in bolting (23.3 ± 0.1 days vs. 23.4 ± 0.2 days, Fig. 4H and Fig. S6), even though the expression levels of these two TREPH1 isoforms were similar in transgenic plants (Fig. S8). Taken together, out touch treatment delays plant flowering, the *treph1-1* (and also *mkk2*) mutant is defective in this touch response. The transgenic wild type TREPH1 protein, but not point-mutated TREPH1^S625A protein, can rescue this defect in response to touch treatment (Fig. 4I). Thus, the S625 amino acid of TREPH1 is required for TREPH1's function in delaying bolting in response to touch, suggesting the possibility that the touch-induced phosphorylation of TREPH1 is an essential step in the touch-response pathway of Arabidopsis.

### Mechano-transcriptome changes in treph1-1 and mkk2

To investigate whether TREPH1 and/or MKK2 may be essential for touch-induced changes in gene expression, we compared touch-regulated transcript accumulation in wild type, *treph1-1,* and *mkk2*. In wild-type plants, we identified 418 genes that were upregulated by touch (fold change ⩾ 2 and probability ⩾ 0.8) and 87 that were downregulated by this treatment (fold change ⩽ -2 and probability ⩾ 0.8). We compared the transcriptomes with previous transcriptomics results, finding that 92% (47 out of 51 genes, Table 1 of [38]) of previously identified touch-regulated genes were again detected by our RNA deep sequencing. By contrast, only 47% of genes upregulated by wounding (76 out of 162, Table S1 of [42]) were identified among these touch-regulated genes.

Using the Type 1 (| Δ log₂Ratio | ⩾ 1) and Type 2 methods of transcriptomic analysis (see Methods) to analyze the data, we identified genes that showed differential touch-responsiveness in the mutants compared to wild type. Genes whose expression was either up- or down-regulated by touch in wild-type plants but less responsive in *treph1-1* were defined as *TREPH1*-dependent genes. The Type 1 and Type 2 methods revealed 92 and 75 *TREPH1*-dependent touch-responsive genes, respectively. Of the 167 touch-responsive genes analyzed, 67 were identified by both the Type 1 and 2 methods (Fig. 5 A and B and Fig. S9 and Dataset S4). Of these 100 unique touch-responsive and *TREPH1*-dependent genes, 86 (86% of 100 genes) had increased expression in *treph1-1,* suggesting that their expression is suppressed by TREPH1 protein during the touch response, whereas 14 touch-responsive genes showed reduced expression in *treph1-1,* indicating a requirement for TREPH1 for appropriate expression of these 14 genes. We subjected several of these genes to RT-qPCR analysis, including *CALMODULIN-LIKE38* (*CML38,* AT1G76650), *ETHYLENE RESPONSE FACTOR11* (*ERF11,* AT1G28370) and *JASMONATE-ZIM-DOMAIN PROTEIN7* (*JAZ7,* AT2G34600) found by the Type 1 method, as well as an unknown touch-inducible gene (*TCH2K1,* AT1G56660) found by the Type 2 method (Fig. 5 C-F). *CML38* was reported as touch-inducible genes in a previous study [38]. *CML39* (AT1G76640), which was also a touch-induced gene in the same study [38] but

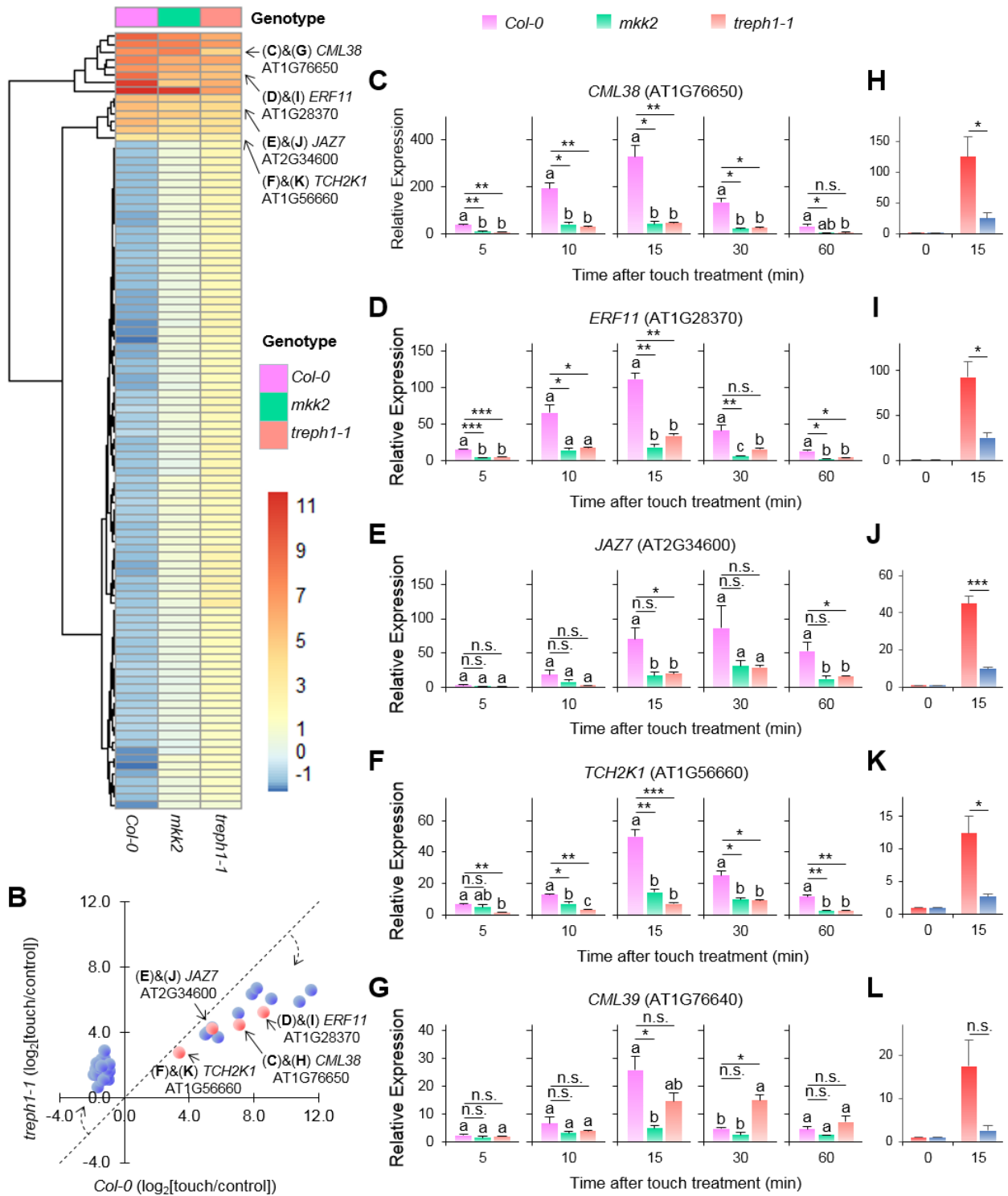

**Figure 5. The roles of TREPH1 and MKK2 in transcriptional regulation during mechanotransduction.**

(*A*) Presentation of the log2 ratios of transcriptomic changes in wild type *Col-0* and the *mkk2* and *treph1-1* mutants after a 10-min time lag following the initial 40-s touch (1 s per touch, Dataset S4). An alternative representation of transcriptomic differences between the wild type and mutant plants is shown in Fig. S9. Only the transcripts showing touch-induced alterations between the wild type and the *treph1-1* and *mkk2* mutants are included in both figures. Arrows indicate the four genes selected for RT-qPCR analysis shown in panels (*C-F*) and (*H-K*). The gradient from blue to red indicate the log2 ratios ranging from -1.9 to 11.5.

(*B*) The log2 ratios of transcriptomic changes in wild type *Col-0* against those in *treph1-1* mutant after a 10-

*(legend continued on next page)*

not TREPH1-regulated according to our transcriptomic results, were also included as controls for touch-inducible genes (Fig. 5G). Although we have detected higher number of genes of increased expression levels in *treph1-1* background (86 out of 100) as compared to that in the wild type plants, no gene was selected from this group for RT-qPCR because most (73%) of these genes are implicated in energy pathways (19 photosynthetic proteins, 13 NADH dehydrogenases, 7 ATP synthases/ATPases, and 1 cytochrome) or protein synthesis (23 ribosomal proteins) (Dataset S4). They are suspected to play an indirect role in plant touch response. We thus decided not to pursue the functions of these genes in touch response in the study.

*CML38* expression was induced 36.7 ± 3.6- and 6.5 ± 0.5-fold at 5 min after the application of a 40-s touch stimulation in the wild type and *treph1-1,* respectively, compared to the control (Fig. 5C). The expression of this gene peaked at a level 328.5 ± 48.8- and 45.3 ± 5.4-fold that of the control at 15 min after touch induction and decreased to 31.7 ± 9.6- and 5.5 ± 1.0-fold at 60 min after touch induction in the wild type and *treph1-1*, respectively. The expression pattern of *ERF11* was similar to that of *CML38,* with its expression level increasing 111.3 ± 7.6- and 33.7 ± 2.5-fold in the wild type and *treph1-1*, respectively, at 15 min after touch induction (Fig. 5D). The other two touch-inducible genes, *JAZ7* and *TCH2K1,* showed similar induction kinetics within 60 min following 40-s of touch treatment in the RT-qPCR-based touch response assay (Fig. 5 E and F). On the other hand, the wild type and *treph1-1* mutant gave no significant difference on the expression of *CML39* at 15 min after the touch stimulation, when the transcriptomic study was performed, although they did show distinct regulation on the expression of *CML39* (4.6 ± 0.6 fold vs. 14.8 ± 1.9 fold) at 30 min after the treatment (Fig. 5G). Furthermore, among these four genes selected based on the results of the *treph1-1* mutant, the *mkk2* mutation also suppressed the expression of three touch-inducible genes *CML38, ERF11* and *TCH2K1* upon touch treatment (Fig. 5 C-G).

To determine whether S625 phosphorylation of TREPH1 is required in the expression of touch-regulated genes, we performed RT-qPCR analysis in transgenic populations of

---

min time lag following the initial 40-s touch (1 s per touch) are shown. The two columns of data for wild type and *treph1-1* shown in (*A*) are plotted. The dashed straight line is the bisector line of the quadrants. The dashed arrows indicate the shift of log2 ratios between wild type and *treph1-1*. The red nodes are the four genes selected for RT-qPCR analysis shown in panels (*C-F*) and (*H-K*).

(*C-G*) The expression of *CML38* (AT1G76650, *C*), *ERF11* (AT1G28370, *D*), *JAZ7* (AT2G34600, *E*), *TCH2K1* (AT1G56660, *F*), and *CML39* (AT1G76640, *G*) is induced within 1 hour after 40-s cotton swab touch treatment in *Col-0, mkk2*, and *treph1-1* plants. The first four genes (*C-F*) were selected based on the results of the transcriptomic analysis (*A* and *B*), while the last (*G*) was selected based on a previous report (Lee et al., 2005).

(*H-L*) Changes in mRNA abundance of *CML38* (*H*), *ERF11* (*I*), *JAZ7* (*J*), *TCH2K1* (*K*), and *CML39* (*L*) induced by 40-s cotton swab touch treatment in TREPH1/*treph1-1* and TREPH1$^{S625A}$/*treph1-1* transgenic plants.

The mRNA levels were quantified using RT-qPCR (see Methods). Means ± SE are shown (three biological replicates). Statistical analysis was performed by Student's *t*-test and Tukey's range test. Significance at *p < 0.05, p < 0.01, p < 0.001*, and *p ≥ 0.05* for pairwise *t*-test are shown as *, **, ***, and n.s., respectively. The homogeneity of variance in each panel with multiple (more than two) values was analyzed using Tukey's range test. Different letters represent significant differences at the 5% level.

*dPro35S::His::YFP::TREPH1/treph1-1* and *dPro35S::His::YFP::TREPH1^{S625A}/treph1-1* (Fig. 5 H-L). Although the presence of the *dPro35S::His::YFP::TREPH1* transgene rescues the *treph1-1* touch expression inducibility of the genes tested, the presence of the *dPro35S::His::YFP::TREPH1^{S625A}* in the *treph1-1* background resulted in reduced expression levels of *CML38, ERF11, JAZ7, and TCH2K1* but not *CML39* (Fig. 5 H-L). These results demonstrate a critical role for Serine 625 in the functioning of the TREPH1 protein, and suggests that S625 phosphorylation upon touch may play a critical role in touch-related gene expression.

### *Biochemical, physiological and homology-based structural analysis of TREPH1*

BLAST analysis of the amino acid sequence of TREPH1 against the proteomes of both plants and animals revealed homologs of the TREPH1 C-terminus in mammals, monocots, and dicots (Fig. 6A). Homology-based tertiary structure (3D) prediction (Fig. 6B) also showed that TREPH1 protein may consist of five tandemly arranged three-helix coiled-coil domains, have a sickle-shaped structure, and lack a transmembrane domain. In contrast, the TREPH1 protein was predicted to present in the sucrose-isolated membrane protein fraction [43] even though there is no trans-membrane domain present in the TREPH1 protein. To confirm if it is a membrane protein, we fractionated TREPH1 protein into both membrane and soluble fractions (see Methods for detail), followed by immunoblot analysis using both anti-S625-non-phosphorylated antibody (anti-npTREPH1) and anti-S625-phosphorylated antibody (anti-pTREPH1) (Fig. 6C). The result showed that the TREPH1 protein is indeed soluble and not integral to microsomal membranes. The molecular biology data deposited in ePlant (http://bar.utoronto.ca/eplant/) also indicates that *TREPH1* gene expression is constitutive in the life cycle of plants.

To investigate how TREPH1 might function in the touch response, we analyzed the interactomics of the phosphoproteins listed in Table 1 using the bioinformatics program STRING. Interestingly, we identified two protein-protein interactomic modules among these touch-related phosphoproteins (Fig. 6D). One module resembles the mechano-signalosomes found in animal cells [6], consisting of dual-tethered architectural integrins that span the plasma membrane and physically interconnect the intracellular actomyosin cytoskeleton with extracellular matrix architectural fibronectin proteins. Another module is mainly composed of kinases and phosphatases of the MAPK cascade interconnected with the $Ca^{2+}$-dependent kinase-mediated phosphor-relay pathway (Fig. 6D). Since plant cells generate cytoplasmic calcium spikes in response to touch [37], leading to touch-inducible gene expression [2], we measured the luminescence emitted from BRET-based *GFP::Aequorin* fusion protein (G5A) [44] on both *G5A/Col-0* and *G5A/treph1-1* transgenic plants following touch treatments. The cytoplasmic calcium signals became stronger as the duration of touch treatment increased (Fig. 6E). No significant difference in calcium signature was observed between plants in the *Col-0* versus *treph1-1* background in response to 1-, 10-, 20-, 40-, and 60-s touch treatments (Fig. 6E), indicating that TREPH1 protein does not function at the upstream of cellular calcium signal. In *Col-0,* the signal strength appeared to be saturated after 40 s. These results are consistent with the dose-dependent morphological changes described above (Fig. 1A). As shown in the study of cellular calcium signals, TREPH1 phosphorylation functions downstream of or in parallel with calcium signaling. As a cytoplasmic and soluble protein that localizes nearby plasma membrane, TREPH1 might function to perceive mechanical signals from hypothetic force receptors instead of calcium ions, then transduce them to the downstream regulation of touch-inducible gene expression, and finally lead to thigmomorphogenesis (Fig. S10).

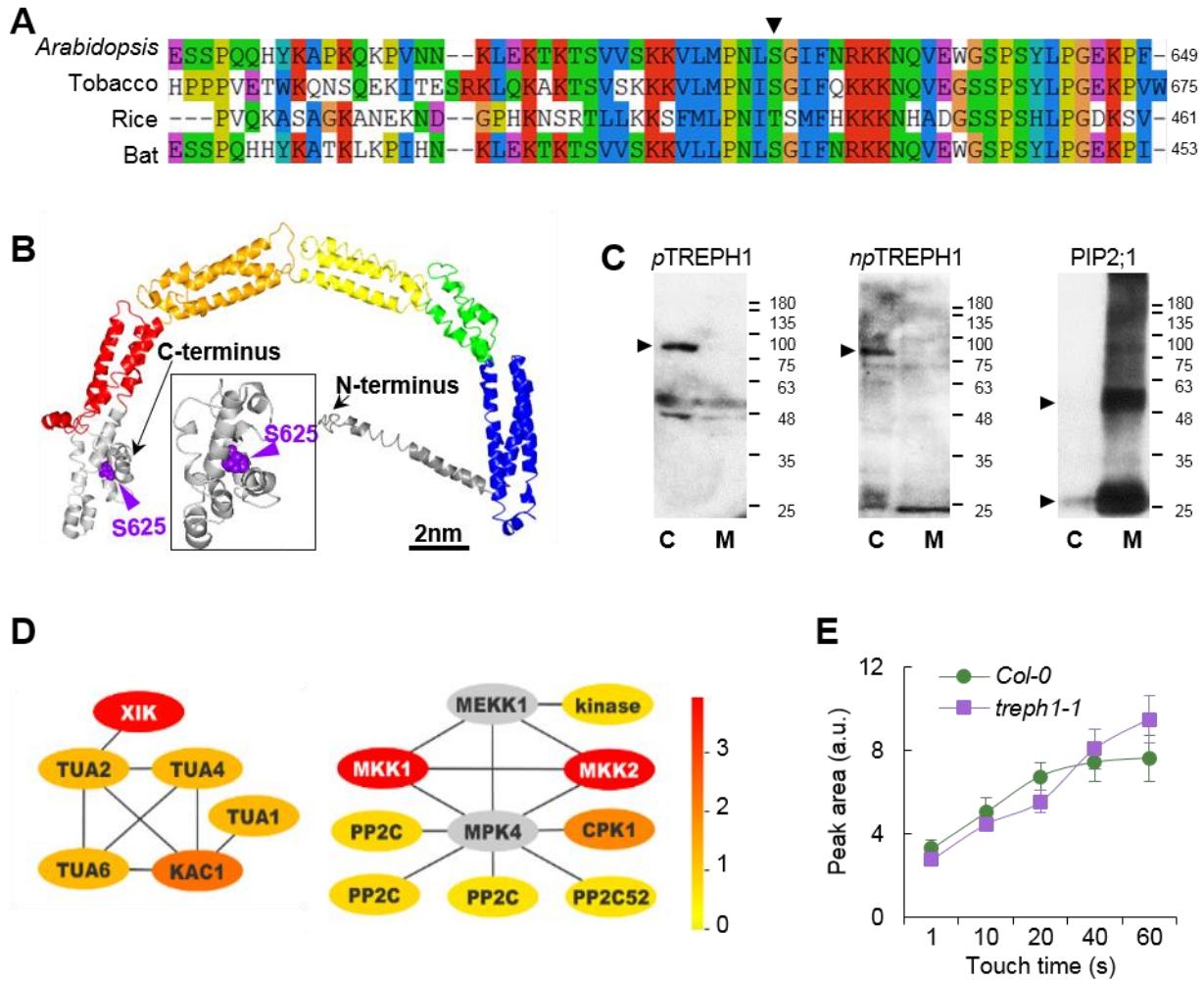

**Figure 6. Bioinformatic and biochemical analysis of the key roles of TREPH1 in mechanotransduction.**

(*A*) Conserved C-terminal sequences of TREPH1-like protein across species. *TREPH1*-like genes are present in the genomes of various plant and animal species. The C-termini of TREPH1-like proteins from tobacco (*Nicotiana tabacum*, XP_016476003), rice (*Oryza sativa*, Os12g0271600), and bat (*Rhinolophus sinicus*, XP_019577179) were aligned with that of TREPH1 protein. Triangles (▼) indicate phosphorylation sites.

(*B*) Homology-based prediction of the molecular structure of TREPH1.

(*C*) Biochemical analysis of cytosolic and integral membrane proteins in the wild type plant cells. The results from left to right was anti-*p*TREPH1, anti-*np*TREPH1 and anti-PIP2;1 immunoblotting, respectively. C and M stand for the cytosolic and integral membrane protein fractions, respectively. Arrows in the left flank indicate the target protein bands. Antibody specificity is shown in Fig. S3 a and h.

(*D*) Model molecular network assembled based on information from the quantitative phosphoproteomic analysis. The touch-induced protein phosphorylation level alterations obtained from quantitative phosphoproteomics (Table 1) are represented in a red (higher PTM level) to yellow (lower PTM level) gradient, where gray is unknown. Node kinase, Protein kinase family protein AT5G09890; three PP2C nodes, PP2C family proteins AT1G16220, AT1G79630, and AT1G03590. Mechanotransduction modeling based on bioinformatic analysis were shown in Fig. S10.

(*E*) The relative peak areas of bioluminescence emitted from the recombinant G5A protein upon various lengths (in seconds) of touch treatments. Means ± SE are shown (*n* ≥ 10, n = number of individuals).

## Discussion

The machinery responsible for mechanosensitivity, including *bona fide* mechanosensitive channels and multimeric mechano-signalosomes, is believed to have evolved multiple times during flowering plant evolution [22,45]. Numerous force sensors may coordinate multiple levels of mechanical signaling originating from intracellular molecular activities and extracellular mechanical signals during versatile plant mechano-responses [46]. One group of sensors includes the mechanosensitive divalent and monovalent ion channels [5], such as MSL, MCA, Piezo, and TPK, and the other group comprises membrane-bound receptor-like Wall-Associated Kinases (WAKs) and other Receptor-Like Kinases (RLKs) [47-49]. These kinases span the plasma membrane, with one terminal domain linked to the extracellular cell wall and the other terminal domain harboring a cytoplasmic kinase. RLKs are thought to function as kinase activity-associated mechanosensors that regulate downstream calcium transients or directly transduce a mechanical phosphor-relay to nuclear events (Fig. S10) [47]. The mechanical signal-induced $[Ca^{2+}]_{cyt}$ signature has a time scale ranging from 5 s to several minutes [37,50]. CDPKs and Calcium Binding Proteins (CBPs) are thought to respond to force signals at the time of the initial touch treatment [51], and these CDPKs might trigger protein phosphorylation within seconds. It is therefore plausible that protein phosphorylation, one of the most abundant post-translational modifications [52], may act to transduce mechano-signals through ion channel-dependent and/or -independent fashion. Our discovery of 24 rapidly mechano-responsive phosphosites of protein groups (Table 1) supports this view.

Like membrane-bound mechano-signalosomes found in animals [6,35], kinases may be part of the force-sensing signalosomes in plants and may span the plasma membrane and be tethered to the extracellular matrix or cell wall and to the, cytoskeleton, enabling perception of deformation and plasma membrane tension imposed by mechanical forces such as touch. For example, a 500-kDa WAK1 protein complex contains a WAK1 kinase subunit, a glycine-rich extracellular protein subunit, an AtGRP-3, and a cytoplasmic type 2C protein phosphatase, KAPP, which can associate with a Rho-related protein [53-55]. Interestingly, we also found that kinases/phosphatases MKK1/2, CPK1, CPK5/6, and four PP2C family proteins, as well as the cytoskeletal protein XIK, kinesin-like (KAC1), and tubulin (TUA1/2/4/6), were among the rapidly and highly phosphorylated proteins identified after a 40-s force-loading treatment (Table 1 and Fig. 6D). These rapid changes upon mechanostimulation suggest the possibility that these phosphoproteins may play roles in early force signaling in plant cells.

The phosphor-relay-mediated cell signaling components, MAP kinases and calcium-dependent protein kinases (Table 1), are involved in various abiotic and biotic stress signaling pathways [56]. Our quantitative phosphoproteomic study indeed found the touch-induced phosphopeptides shared by MKK1 and MKK2 phosphoproteins. These two kinases have overlapped functions in some cases [57,58]. For example, MKK1 and MKK2 are both implicated in hormone-mediated biotic stress responses [57,59]. However, different kinase activities of these two MKKs were also detected under various stress conditions [57,60]. The amino acid sequence alignment between MKK1 and MKK2 proteins showed that the identity and similarity of these two proteins were only 57.3% and 69.7%, respectively. Given the possible presence of multiple pathways in mechano-signaling, both MKK1 and MKK2 kinases may mediate the force signaling through identical and/or distinct pathways, and they probably function in different cellular events during touch response and respond to different strengths of mechanical stimulation. One example used to justify such a speculation is the case of cold stress signaling. It has been reported that 10 min of 4°C chilling treatment of plants triggered the activity of MKK1 in phosphorylating MPK4 [61], whereas 10 min of 0°C freezing treatment only induced the kinase activity of MKK2 rather than MKK1 [60]. Thus, MKK1 and MKK2 may be able to transduce cell signaling *via* multiple pathways. Taken together, these intricate results suggest that MKK2 and MKK1 may transduce force signals through an overlapping and/or kinase-specific pathways (either MKK1- or MKK2-dependent pathway), and their functions may be both additive and synergistic in one or more aspects of the touch response.

The 3C quantitative PTM proteomic workflow generally produces a series of candidate proteins with an arbitrarily defined confidence level (for example, 95% of positive rate, as used to select the touch-regulated phosphosites shown in Table 1). As a result, an acceptable percentage of false positive are usually present among the selected phosphoproteins (BH-FDR is equal to or smaller than 5%). To elucidate the function of these phosphorylation modifications selected by a statistical evaluation, it is important to apply either an immunoblot assay or an additional genetic screen or both to further pinpoint those candidates of key functions in touch response with a minimum effort. The functional studies, such as physiological and molecular genetic characterization, are believed to complement the Omics approach and viewed as the second dimension of confirmation in the 3C quantitative PTM proteomic workflow. In this real case of proteomic study, the application of Benjamini-Hochberg hypothesis test correction has allowed us to select 24 touch-regulated phosphosites with 5% BH-FDR cutoff in order to save labor and time used for genetic screening. Given that the probability of finding at least 4 successes out of 5 randomly selected candidates is over 99%, 5 T-DNA insertional mutants were considered to provide us at least 4 opportunities to find a touch response mutant if the loss of any of these phosphosites has functional effects on the touch-induced morphogenesis. Otherwise, we would have to screen at least 10 T-DNA insertional mutants to reach the same confidence level if we have used the results based on the *t*-test ($p < 5\%$, Dataset S3). After a comprehensive functional analysis of TREPH1, several lines of evidence suggest that TREPH1 is involved in one aspect of the touch response: first, TREPH1 was rapidly phosphorylated at S625 (to 7.3-fold control levels) upon 40-s of touch treatment; second, the T-DNA insertional mutant *treph1-1* exhibited an abnormal touch response, failing to show a delay in flowering time (Fig. 4 and Fig. S6); third, a non-phosphorylatable TREPH1 isoform fusion (TREPH1$^{S625A}$) failed to complement the *treph1-1* mutant (Fig. 4 and Fig. S6).

Modeling suggests that TREPH1 may possess a sickle-shaped structure of five tandemly arranged three-helix coiled-coil domains and lacks transmembrane domain (Fig. 6B). Intriguingly, the putative TREPH1 structure resembles the cytoplasmic domain of bacterial cell division protein EzrA, which regulates the formation of the cytokinetic Z-ring formed by the tubulin homolog FtsZ protein [62], suggesting that the TREPH1 may interact with tubulin and FtsZ-like protein in plant cells.

A dynamic change at the gene expression level is associated with the morphological changes that occur during the plant touch response [2,4,63]. Many genes are transiently altered in expression, with transcript levels changing within 5 min upon touch treatment, peaking between 10 to 30 min, and returning to basal levels within one to two hours (Fig. 5) [64,65]. In the present study, we found that the touch-triggered phosphorylation events occurred within 40 seconds and coincided with luminescence reporting of cytoplasmic calcium ion levels. TREPH1 is not required for touch-induced calcium signaling changes; fluorescence reporting of intracellular calcium levels show clear responses to 10- to 60-s touch treatments in *treph1-1*, comparable to that of wild type (Fig. 6E). These results are consistent with TREPH1 acting downstream of or in parallel with calcium signaling.

To integrate our results, we developed a force signaling model for the Arabidopsis touch response, shown in Fig. S10. One possibility is that the cytoplasmic and soluble TREPH1 protein may associate with mechano-signalosomes or force-activated ion channels to transduce the touch signal downstream. MAP kinases undergo crosstalk with calcium-dependent protein kinases during abiotic and biotic stress signaling [56]. One possibility is that these two touch-signaling pathways regulate the expression of the same group of genes, which are regulated during plant growth and bolting [66,67]. Even though the substrates of the touch-responsive MKK2 and the CPK5/6 and CPK1 kinases are not known, ERF11 and JAZ7 transcription factors may serve as either direct or indirect substrates of these phosphorylation/dephosphorylation enzymes. ERF11 functions in gibberellin (GA) biosynthesis, GA-regulated flowering, and ethylene biosynthesis [68]. The phosphorylated isoform of ERF110 promotes bolting in Arabidopsis, while the overexpression of S62-non-phosphorylated ERF110 suppresses bolting [69]. On the other hand, JAZ7 regulates the jasmonic acid response and flowering in Arabidopsis [70]. Taken together, we hypothesize that rapid touch-responsive kinases and phosphatases quickly modify both the expression level and PTM code (including phosphorylation) of these regulators to regulate the touch response. In the *treph1-1* and *mkk2* mutant genetic backgrounds, the expression of these downstream bolting regulatory genes is suppressed (Fig. 5 and Fig. S9), leading to an early bolting phenotype (Fig. 4 and Fig. S6 and S7).

In conclusion, protein phosphorylation is a rapid and broad response to touch stimulation in Arabidopsis and may play a critical role in the mechanoresponse pathway of plants.

## Materials and Methods

### *Mutant screening and transgenic plants*

For the mutant screening, the T-DNA left border (LBb1) primer, along with gene-specific left or right primer (LP/RP), was used to confirm the insertions. All the primers used to screen homozygous plants were listed in Table S1.

For the transgenic plants, the *ProCML39::LUC/Col-0* constructs were generated using the Gateway-compatible binary vector pBGWL7 [72]. The 2 Kb region upstream of the *CML39* start codon was PCR-amplified using the following primers (Table S1):

LUF: 5′-CACCAAACTTTGCCGGAAACTATCAC-3′;

LUR: 5′-ATACCCGGGTTTGAGAAAGAAAAGATTGTATTTG-3′.

The amplified product was cloned into the pENTR/D-TOPO entry vector (Invitrogen) according to the manufacturer's instructions. The authenticity of the cloned DNA insert was confirmed by endonuclease restriction digestion and DNA sequencing. The pENTR/D-TOPO entry clones containing the promoter region of *CML39* were subsequently ligated to pBGWL7 and recombined using the LR Clonase reaction (Invitrogen) according to the manufacturer's instructions. Positive clones were confirmed by endonuclease restriction digestion analysis and DNA sequencing prior to being transformed into *Agrobacterium*.

Both the wild-type and Ser625Ala (S625A) recombinant *TREPH1* genes were generated as described [73]. A 2.0 kb DNA fragment encoding the full-length TREPH1 was amplified from Arabidopsis genomic DNA by PCR using the following primers (Table S1):

CPF, 5'-TCAT<u>GGCGCGCC</u>ATGGTTGCTAAGAAGGGACGTAG-3' (*AscI* site underlined);

CPR, 5'-GGG<u>GAGCTC</u>AAAAGGGTTTCTCTCCAGG-3' (*SacI* site underlined).

Point-mutated *TREPH1^S625A* was generated by PCR using mutagenic primers (mutation sites are indicated in bold):

PPF, 5'-CCGAATCTA**GC**TGGAATCTTC-3' (mutation sites bolded);

PPR, 5'-GTTGAAGATTCCA**GC**TAGATTCG-3' (mutation sites bolded).

The resulting recombinant binary vectors harboring the *His::YFP::TREPH1* and *His::YFP::TREPH1^S625A* fusion genes (pIYFP, AY653732) [74] were placed under the control of a double cauliflower mosaic virus 35S promoter which contains -829 to -62 and -829 to -1 with TATA box regions (construct pHub10) [75], respectively.

All transgenic plants, *ProCML39::LUC/ Col-0*, *dPro35S::His::YFP::TREPH1/treph1-1*, and *dPro35S::His::YFP::TREPH1^S625A/treph1-1* were generated using the floral dip method [71].

The shorter version name, *TREPH1/ treph1-1* and *TREPH1^S625A/ treph1-1*, used in the manuscript annotates transgenic plants *dPro35S::His::YFP::TREPH1/ treph1-1* and *dPro35S::His::YFP::TREPH1^S625A/ treph1-1*, respectively.

### Plant growth

Arabidopsis seeds were sown on standard or [15]N-enriched Stable Isotope Labeling in Arabidopsis (SILIA) solid agar medium [36] in 7.7-cm diameter, 12.7-cm high glass jars placed in growth chambers under a 14-h-light/10-h-dark regime. All the jars were covered with Hydrophobic Fluoropore Membrane (http://www.shjiafeng.com, Jiafeng, Shanghai, China) to keep the growth condition air permeable and sterile. Each jar contained 30-40 ml medium and 3 plants were grown in a jar to minimize contact between plants as well as in between plants and glass wall. For soil-grown plants, soils from Jiffy Products International BV (The Netherlands, 1000682050) and from Plantmate (Hong Kong, 65310) were mixed together at a ratio of 2:1. Then soil-grown plants were grouped into big trays and placed in a growth room under a constant light regime. Soil-grown plants were irrigated every two days with 1.5 L of water after germination. The light intensity applied to both touch and untouched plants was 180 – 240 µE m$^{-2}$ s$^{-1}$, which was measured by an IL 1700 research radiometer (International Light, Newburyport, MA). In both growth chambers and rooms, the temperature was set at 23.5 ± 1°C, and humidity was 35%-45%.

### Force treatment

Thigmomorphogenetic changes under long-term mechanical stimulation were examined in wild-type Arabidopsis ecotype *Col-0* as well as *treph1-1*, *mkk2*, other T-DNA insertional mutants, and *LUC, TREPH1/treph1-1*, and *TREPH1$^{S625A}$/treph1-1* transgenic plants. To observe morphological touch responses (thigmomorphogenetic changes), 12–14 day-old plants grown either in solid agar medium or in soil under a constant light condition (180 - 240 µE m$^{-2}$ s$^{-1}$) were subjected to long-term mechano-stimulation, which lasted for additional 18–25 days until all plants bolted, depending on the genotypes of Arabidopsis plants and the growth conditions. Cotton swab (Video S1 and S2) and finger touch treatments were performed 3–4 rounds daily, with 10 – 80 touches (one touch per second) every round. Human hair was also used for touch treatment (Video S3). Hair touch was controlled by a fully automated in-house-built machine (HKUST Laboratory Service). The hair touch treatment was performed 3 rounds per day with 8 hours long interval. In each round, the individual plant was touched 40 times like the case of cotton swab touch. Plants were considered to have bolted when the primary inflorescence height reached to 1 cm [2,69]. The percentage of bolting plants of the *i*-th day after sowing was calculated as the numbers bolting plants on or before that day over the total number of plants, in which the data from three biological replicates were pooled together. For the transgenic plants, the bolting data of individuals from seven and six independent transgenic lines expressing *His::YFP::TREPH1* and *His::YFP::TREPH1$^{S625A}$* proteins, respectively, were pooled together to calculate the quantitative bolting phenotype and to perform the statistical evaluation on the bolting curve as described in other reports [76,77].

Short-term mechanical force stimulation was performed using three techniques: touch, air blowing (to simulate wind), and water sprinkling treatment (to simulate rainfall). Based on the touch strength applied in a previous experiment, i.e., approximately 4 g/cm$^2$ [64], for both cotton swab and finger touch treatment, a 1 mN ( ≈ 4 g/cm$^2$ × 10 m/s$^2$ × 0.1$^2$π cm$^2$) force was applied to one small circle (0.1 cm radius) at a time. The short-term touch treatment was applied for 10 to 150 seconds at a frequency of one touch per second. Thus, the accumulated force applied to one plant was 10 mN (10 times), 20 mN (20 times), 40 mN (40 times), 60 mN (60 times), 80 mN (80 times), or 150 mN (150 times). Both wind and rain – mimicking stimuli were applied to the plants at the same level of strength. These force treatments were applied to 3-week-old plants grown on agar (Video S1) or in soil (Video S2 and S3) for various lengths of times; the plants were then subjected to protein phosphorylation, gene expression, and/or cytoplasmic calcium flux measurements. Both the untouched control and touch-treated plant tissues (aerial portion) were immediately frozen in liquid nitrogen (within 1 second of treatment) and stored at -80°C for subsequent analysis.

### Isolation of 14N/15N-labeled total cellular proteins and in-solution protease digestion

Control and touch-treated Arabidopsis plants were labeled with $^{15}$N and $^{14}$N stable isotopes following the established SILIA method (Fig. S1) [36,39]. Tissues from touch-treated ($^{14}$N-labeled) and control ($^{15}$N-labeled) plants were combined to form the Forward mixing replicate, and touch-treated ($^{15}$N-labeled) and control ($^{14}$N-labeled) samples were combined to form the Reciprocal mixing replicate. Distinct biological replicates used plants samples that were prepared independently during the whole process, including seeds preparation, medium preparation, growth, harvest and the following biochemistry experiments. Different biological replicates were prepared and performed at different times by the same person or different persons at the same time. Because plants in both $^{15}$N- and $^{14}$N-coded media were grown under separate and distinct biological growth conditions, each Forward and Reciprocal mixing replicate was classified as a separate biological replicate. In total, six mixing replicates were performed, which are equivalent to six biological replicates. Plants grown in 80 – 120 transparent glass jars (including both $^{14}$N/$^{15}$N-labeled plants at a ratio of 1 : 1) were harvested for each biological replicate, and the weight of mixed aerial tissue was 30 – 40 g. $^{14}$N/$^{15}$N-labeled total cellular protein isolation was performed under fully denaturing conditions beginning when the frozen tissues were thawed in urea extraction buffer (UEB). The protein of each biological replicate weighed 0.3 – 0.5 g, and the yield was 1% - 1.25%. Peptides were prepared according to an established protocol [43,78] and 150 – 250 mg peptide (yield, about 50%) was used for each biological replicate.

### TiO2/Fe3+-IMAC phosphopeptide enrichment and HPLC fractionation

To increase the ratio of phosphopeptides to total peptides, an integrated multi-step and tandem affinity column enrichment protocol was developed to enrich for phosphopeptides [78]. This modified enrichment method included the use of both TiO$_2$ and Fe$^{3+}$-NTA IMAC beads; enrichment was performed multiple times. The purity of the phosphopeptides normally reached >80%.

The highly purified phosphopeptides were fractionated into 15 – 20 fractions on a weak anion-exchange (WAX) column (PolyLC Inc., Columbia, MD, USA) *via* HPLC. The Poly WAX LP$^{TM}$ Column (Item 104WX0503) was 100 × 4.6 mm in size, with a 5 μm particle diameter and a 300 Å pore diameter. HPLC was performed as previously described [79].

### LC-MS/MS analysis

Phosphopeptides prepared from two forward replicates were analyzed on a nano-Acquity system (Waters) connected to an LTQ-Orbitrap XL hybrid MS (Thermo Fisher Scientific, San Jose, CA, USA), while the two remaining forward and two reciprocal replicates were analyzed on a Q-Exactive hybrid quadrupole-Orbitrap MS (Thermo Fisher Scientific), interfaced with a nano-Ultra performance liquid chromatography system (Easy nLC, Thermo Fisher Scientific).

For the two forward replicates analyzed by LTQ-Orbitrap XL hybrid MS, the phosphopeptides were separated on a gradient of 2% to 98% solvent B (acetonitrile with 0.1% (v/v) formic acid) for 240 min. Solvent A comprised 0.1% (v/v) formic acid in water. The mass spectrometer was operated in positive ion mode with the following basic parameters: the MS survey scan was performed in a Fourier transform cell with a window between 350 and 1,800 m/z. The resolution was set to 60,000. The m/z values triggering MS/MS were placed on an exclusion list for 60 s. The minimum MS signal for triggering MS/MS was set to 5,000. There were 11 scanning events. The CID acquisition method was performed with a target value of 5,000 in the linear ion trap, collision energy of 35%, Q value of 0.25, and activation time of 30 ms.

For the four remaining replicates (two forward and two reciprocal) analyzed by Q-Exactive hybrid quadrupole-Orbitrap MS, the phosphopeptides were separated on a gradient of 3 to 85% buffer B with a 500 nl/min flow rate for 300 min. The mass spectrometer was operated in positive ion mode with the following basic parameters: MS survey-scan range of 300 and 2,000 m/z. The resolution was set to 70,000 with an AGC target value of 1e6 ions and a maximum ion injection time of 60 ms. Resolution of dd-MS2 was set to 17,500 with an AGC target value of 1e5 ions and a maximum ion injection time of 100 ms. The stepped NCE was 27 with a 10 TopN.

A monolithic silica analytical column (Monocap C18, High Resolution 2000, GL Sciences Inc., Tokyo, Japan) with an internal diameter of 0.1 mm, a length of 2,000 mm, and a pore size of 2 μm was used to analyze all six replicates.

### *Quantitative phosphoproteomic analysis*

The resulting LC-MS/MS raw data were converted to both mzXML and mgf formats following an established procedure [39]. All MS2 spectra contained in the mgf files were searched against TAIR10 (35,386 proteins) using the Mascot search engine (Version 2.3, Matrix Science) as described [73]. The false discovery rate (*FDR*) defined to select the PSM of phosphopeptides was set to 1% [80]. Mascot delta score of 10 was employed as the cutoff for reliable phosphosite identification, which roughly corresponds to a ~1% false localization rate (*FLR*) according to a previous study [81], indicating that the probability that a site is correct is 10-times higher than that for other possible sites (http://www.matrixscience.com). Phosphopeptides were selected for quantitation according to the following criteria: i) present in both Forward and Reciprocal replicates, ii) found in at least three biological replicates, iii) present in a pair (both $^{14}$N- and $^{15}$N-coded) from at least two replicates, and iv) have at least five MS2 spectra (four PSM out of five from the two pairs of MS1 isotopologs). Ion chromatograms (MS1 ion intensities) from both $^{14}$N- and $^{15}$N-coded isotopic envelopes of all selected phosphopeptides were quantified using both the in-house built Siliamass program and the software Silique-N (Alpharomics Co., Limited, Shenzhen, China, http://www.alpharomics.com/Home/Product_msda). Phosphopeptides with at least five ratios of light/heavy MS$^1$ isotopologs and at least one-third of the ratios coming from either Forward or Reciprocal replicates were included in the final quantitative analysis (Dataset S3). Finally, the statistical significance of the ratios was determined using a two-tailed Student *t*-test, followed by Benjamini-Hochberg multiple hypothesis testing correction (BH-FDR $\leqslant$ 5%) [82].

### *Bioinformatics analysis of phosphosites*

Both the chosen touch-regulated phosphopeptides (Table 1) and the bioinformatics BLAST-based putative phosphosites (Fig. 2 and Fig. S2) were included in phosphorylation site motif construction. Standard phosphopeptides of equal length (±6 amino acids surrounding a phosphosite) for all experimentally identified phosphosites were classified with Motif-X [83], followed by MEME (Multiple Em for Motif Elicitation) [84], to identify frequently detected motifs around the phosphosites. Default parameters were used for MEME, except that the parameter "minimum sites for MEME" for each motif was changed to 6. The resulting motifs were adjusted manually. The peptides were removed if their phosphorylated residues were aligned to another position, and entries of low conservation at highly conserved residues (i.e., <2 bits) were also removed. As a result, the peptides of higher conservation ($\geqslant$2 bits) were defined as conserved phosphosite motifs. BLAST-based touch-regulated phosphorylation motif construction was performed for all touch-regulated phosphopeptides (Fig. 2 and Table 1 and Fig. S2) as described previously [78]. The Logo figures were generated using WebLogo (http://weblogo.berkeley.edu/logo.cgi) [85].

### Antibody preparation and immunoblot assays

Rabbit polyclonal antibodies raised against the S625-phosphorylated peptide of TREPH1 and the T29-phosphorylated peptide of MKK1 / T31-phosphorylated peptide of MKK2 were produced commercially (GL Biochem Ltd. Shanghai, China). The synthetic oligopeptide pTREPH1 and pMKK1/2, which were used to produce the polyclonal antibodies, were [619]VLMPNL*pS*GIFNR and [25]FLTQSG*pT*FKDGDL, respectively. Rabbit polyclonal anti-GFP antibody (sc-8334) and anti-Actin antibody (sc-1615) were purchased from Santa Cruz Biotechnology (Santa Cruz, CA, USA). Plant protein was prepared for immunoblot analysis as previously [73]. Total cellular proteins were extracted under fully denaturing conditions with UEB. The proteins were fractionated on 8% or 10% SDS-PAGE gels and immobilized onto PVDF membranes (GE Healthcare). The immobilized membranes were blocked with 5% (w/v) milk in TBST (Tris-buffered saline, 0.1% (v/v) Tween 20) for 1 h at room temperature, followed by incubation with primary antibody for 1 h at room temperature and three washes (10 min each) with TBST. The membranes were then incubated with secondary antibody for 1 h at room temperature (anti-rabbit, conjugated with HRP, Bio-Rad), followed by three washes (10 min each) with TBST. Luminata™ Forte Western HRP Substrate (Millipore) was used for membrane exposure and signal detection. Immunoblot quantification was performed exactly as described (http://www.lukemiller.org/journal/2007/08/quantifying-western-blots-without.html).

### Microsomal protein and cytosolic protein extraction for subcellular fractionation

Microsomal membrane protein and cytosolic protein extraction were performed as described [73], with several modifications. The frozen tissue was lysed with microsomal isolation buffer containing 150 mM Tris–HCl (pH 7.6), 8 M urea, 5 mM DTT, 5 mM ascorbic acid, 20 mM EGTA, 20 mM EDTA, 50 mM NaF, 1 mM PMSF, 1% (w/v) glycerol 2-phosphate, 0.5% (v/v) phosphatase inhibitor cocktail 2 (Sigma), EDTA-free protease inhibitors cocktail (Complete™, Roche), and 2% (w/v) polyvinylpolypyrrolidone. Cell debris was removed by 10,000 x g centrifugation for 20 min at 12°C. The supernatant was centrifuged at 110,000 x g for 2 h to fractionate the microsomal membranes from the cytosolic protein present in the supernatant. The microsomal membrane proteins were extracted with the same microsomal protein isolation buffer and centrifuged at 110,000 x g for 2 h to remove the lipids. Both the microsomal protein and cytosolic protein were precipitated twice with nine volumes of pre-cooled (–20°C) trichloroacetic acid/acetone (1:8 v/v) solvent for at least 2 h.

### Transcriptomics and RT-qPCR

Wild-type *Col-0*, *treph1-1*, and *mkk2* tissues were collected three times for mRNA isolation. Control and treated plants were touched with a cotton swab for 40 seconds (1 touch per second) and harvested immediately and 10 min later, respectively. A total of 18 samples were collected for mRNA sequencing. Total RNA was extracted using the CsCl method [86,87], and RNA integrity was confirmed by formalin agarose gel electrophoresis and with a Bioanalyzer 2100 (Agilent Technologies, CA, USA). The cDNA library was prepared and sequenced on the BGISEQ500 platform by BGI (Shenzhen, China). After filtering the low-quality reads, Bowtie2 [88] was used to map clean reads to reference genes (ensembl_release31) and HISAT [89] to the reference genome (TAIR10). The quantification program RSEM [90] was then used to analyze these mapped reads, and the expression level was calculated by the FPKM method. The significance of differential expression was calculated using the NOISeq method [91]. Both fold change $\geqslant 2$ and diverge probability $\geqslant 0.8$ were chosen as the thresholds to define significantly differentially expressed genes. To identify genes whose expression was affected by TREPH1 protein, various approaches were applied. First (Type 1), the threshold

$$\left| \log_2 \frac{E_t^{Col-0}}{E_c^{Col-0}} \right| - \left| \log_2 \frac{E_t^{treph1}}{E_c^{treph1}} \right| \geq 1$$

was applied to identify genes that were significantly regulated by touch treatment in wild type. Here, $E_t^{Col-0}$, $E_c^{Col-0}$, $E_t^{treph1}$, and $E_c^{treph1}$ represent gene expression levels in touch-treated *Col-0*, untouched *Col-0*, touch-treated *treph1-1*, and untouched *treph1-1*, respectively. Second (Type 2), of the genes that showed significantly differential expression levels in touch-treated wild type and *treph1-1* plants, genes that satisfied

$$\left| \log_2 \frac{E_c^{Col-0}}{E_c^{treph1}} \right| < 1 \text{ and } Prob < 0.8$$

or

$$\text{sgn}\left( \log_2 \frac{E_c^{Col-0}}{E_c^{treph1}} \right) = - \text{sgn}\left( \log_2 \frac{E_t^{Col-0}}{E_t^{treph1}} \right)$$

were also selected.

RT-qPCR was performed following a standard protocol. Total RNA was isolated by the TRIzol extraction (Thermo Fisher Scientific) method and treated with DNase to remove DNA contamination. Each RNA sample (0.8 µg) was reverse transcribed using SuperScript III First-Strand (Invitrogen). PCR was conducted on a LightCycler® 480 instrument II (384-well plate, Roche, Basel Switzerland) using LightCycler 480 SYBR Green I Master (Roche) in a 20 µl reaction volume according to the manufacturer's specifications (containing 2 µl 10-fold dilution cDNA reaction mixture, 10 µl 2x SYBR Master Mix, 0.7 µl each of 10 µM gene-specific primer, and 6.6 µl water). The PCR conditions included an initial pre-incubation step for 10 min at 95°C, followed by 45 cycles of denaturation at 95°C for 10 seconds, annealing at 60°C for 15 seconds, and a final extension at 72°C for 15 seconds. The melting curve and cooling steps were followed according to the manufacturer's protocol. Primer3 (http://biotools.umassmed.edu/bioapps/primer3_www.cgi) was used for gene-specific primer design according to a previous study [92]; the primer sequences are listed in Table S1. The Ubiquitin-conjugating enzyme 21 gene (*UBC12*, AT5G25760) used in the previous study [93] was selected as an internal control. The cycle threshold (consistent for each gene within one replicate) and the cycle number (Ct) at that threshold were determined using LightCycler® 480 software. The relative expression level of each gene was calculated by the $2^{-\triangle\triangle Ct}$ method [94].

### Luciferase and Ca2+ imaging

*G5A* transgenic plants were used for $Ca^{2+}$ imaging. 7–14-day-old *G5A* seedlings were incubated in 2.5 µM coelenterazine in the dark for 12 h before being imaged. For the 14-day-old transgenic seedlings expressing *ProCML39::LUC*, luciferin was sprayed on the plants 6 h before observation. The temporal profile of luminescence and fluorescence was detected and recorded using a custom-built photon-multiplier tube (PMT) platform containing a P10PC PMT (Electron Tube Enterprises, Uxbridge, UK) mounted in a dark box (supplied by Science Wares, East Falmouth, MA, USA). Plants were touched with a cotton swab inside the dark box.

### Homology structure modeling

Homology structure modeling was performed by submitting the full-length protein sequence of TREPH1 to the online protein structure predictor I-TASSER (http://zhanglab.ccmb.med.umich.edu/I-TASSER/) [95]. Default settings were used.

### Protein-protein interaction module prediction

All touch-regulated phosphoproteins were submitted to STRING databases (http://string-db.org/) [96] to search for interactions among proteins. No more than five interactions were allowed to be added by the databases to detect indirect links between any two phosphoproteins.

### Statistical analysis

The statistical significance of the results of immunoblotting, morphological analysis, and RT-qPCR was assessed using two-tailed student's *t*-test, with significance represented by *, **, and *** at $p < 0.05$, $p < 0.01$, and $p < 0.001$, respectively. The one-way ANOVA test was performed for multiple comparisons using IBM SPSS Statistics (Version 24, IBM Corporation, Armonk, NY, US). Homogeneous subsets were defined by Tukey's range test at a significance level of 5%. The significance of quantitative RNA sequencing data was calculated with NOISeq [91], and probability > 0.8 was used as a cutoff for touch-inducible genes. Quantitative data are represented as means ± SE.

### Biological Replicates

In this manuscript, biological replicates stand for biologically distinct samples being prepared independently. In the molecular and biochemical experiments, all steps of different biological replicates, including seeds preparation, medium preparation, growth, physiology assay, harvest and the following molecular experiments were performed independently. The samples from different biological replicates were never pooled together at any step. In the morphogenetic experiments, separate biological replicates were done at different times.

### Data availability

## Acknowledgments

This research was supported by grants, 31370315, 31570187 (National Science Foundation of China), 661613, 16101114, 16103615, 16103817, AoE/M-403/16 (RGC of Hong Kong), SRF11EG17PG-A, SRFI11EG17-A (the Energy Institute of HKUST), SBI09/10.EG01-A (the Croucher Foundation CAS-HKUST Joint Laboratory Matching Fund), Rice 04 (the Sponsorship Scheme for Targeted Strategic Partnerships) and GDST16SC02 (the Guangdong-Hong Kong Key Area Breakthrough Program). The authors also want to thank Dr. Marc Knight and Dr. Jean-Baptiste Thibaud for the gift of the Aequorin (AEQ) gene and the GFP-Aequorin fusion gene (G5A), respectively. We also express our gratitude to Dr. Daiying XU and Jingjing Li of NanoBioImaging Ltd., Dr. Yan Zhang of Materials Characterization and Preparation Facility, HKUST, and Dr. Andrew L. Miller and Ms. Mandy Chan of the Division of Life Science, HKUST for their assistance with super-resolution microscopy imaging, scanning electronic microscopy imaging, and for providing us with the custom-built platform used to perform the luminescence $Ca^{2+}$ imaging experiment. The authors also want to express their great thanks to Dr. Junxian He and his PhD student Mr. Wei Yang, for their kindly help with confocal microscopy imaging. Authors would like to thank Nicole Wong and Stacy Zhu for their contribution to the measurement of the touch responses and Jennifer Lockhart and Kathleen Farquharson for editing the manuscript.

## Author contributions

K.W., D.Q, F.R., and Q.Z. performed various mechanical force treatments; K.W. and F.R. performed global phosphopeptides isolation; J.L., W.Z., and C.D. performed preliminary MS/MS analysis; K.W. and Z.Y. performed MS/MS data analysis, cytoplasmic calcium flux measurements, and touch morphological response experiments; K.W., D.Q, and S.L. performed immunoblot analysis; S.L. performed the subcellular protein fractionation; D.Q constructed transgenic *TREPH1/treph1-1* and performed mutagenesis; E.W.C. and J.B. constructed transgenic *LUC* plants; K.W. performed transcriptional study; Z.Y. performed quantitation, bioinformatics and structure modeling; N.L., Z.Y., K.W., J.B., and M.W. wrote the manuscript; N.L. was involved in project planning, experimental design, project execution, supervision of experiments, and communication with collaborators and responsible for distribution of materials integral to the findings.

***Supporting Information (SI)***

Fig. S1. Overview of the stable isotope labeling in Arabidopsis (SILIA)-based quantitative phosphoproteomics.

Fig. S2. Motifs of phosphosites measured by phosphoproteomic analysis of touch-treated and untouched Arabidopsis.

Fig. S3. Immunoblot analysis of phosphopeptides, TREPH1 and MKK1/2 proteins using phosphosite-specific or protein-specific polyclonal antibodies.

Fig. S4. Immunoblot analysis of the force-regulated protein phosphorylation.

Fig. S5. Genotyping and transcript measurement of T-DNA insertional mutants.

Fig. S6. Touch responses of wild type and mutants determined by three biological replicates.

Fig. S7. Morphological measurement of additional Arabidopsis touch-response traits: stem height and rosette size.

Fig. S8. Genotyping of Arabidopsis plants overexpressing TREPH1 and TREPH1$^{S625A}$ protein isoforms.

Fig. S9. Log2-transformed median-centered expression of transcripts showing varied touch-induced alterations between the wild type and *treph1-1*.

Fig. S10. Mechanotransduction modeling.

Table S1. All primers used in this study.

Video S1. Cotton swab touch treatment and harvesting of tissues grown on agar for proteomic analysis.

Video S2. Cotton swab touch treatment of plants grown in soil for biochemical, physiological, morphological and genetic studies.

Video S3. Human hair touch treatment of plants grown in soil for physiological, morphological and genetic studies.

Dataset S1. Phosphopeptides detected repetitively (at least twice) in this study.

Dataset S2. Novel phosphopeptides detected in the present study.

Dataset S3. Quantitative PTM proteomic analysis of the selected phosphopeptides.

Dataset S4. Touch-regulated genes that were differentially regulated in the wild type and *treph1-1* and *mkk2* Arabidopsis.

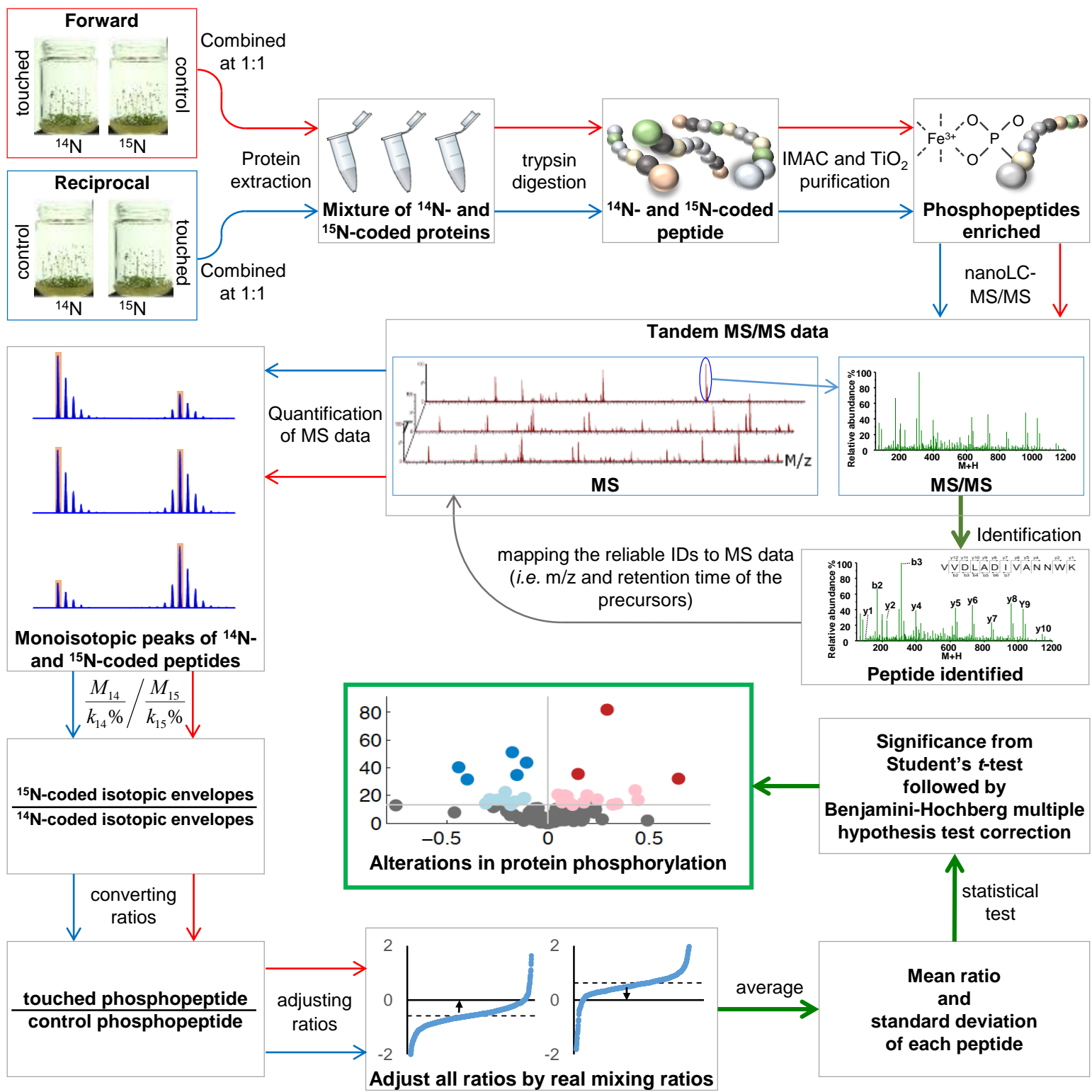

**Fig. S1. Overview of the stable isotope labeling in Arabidopsis (SILIA)-based quantitative phosphoproteomics.** Both forward and reciprocal heavy nitrogen isotope labeling were performed in parallel as shown in red (forward) and blue (reciprocal) arrows, respectively. In the F mixing experiment, the touched and control plants were labeled with $^{14}N$ and $^{15}N$ stable isotope-coded salt, respectively. In the case of R mixing experiment, the two groups of plants were labeled in a reverse order. The data from both F and R mixing experiments were analyzed using the software Silique-N (Alpharomics Co., Limited, Shenzhen, China), in which all data were quantified and put together (green arrows) to infer the final results. It is related to Fig. 2 and Dataset S1-S3.

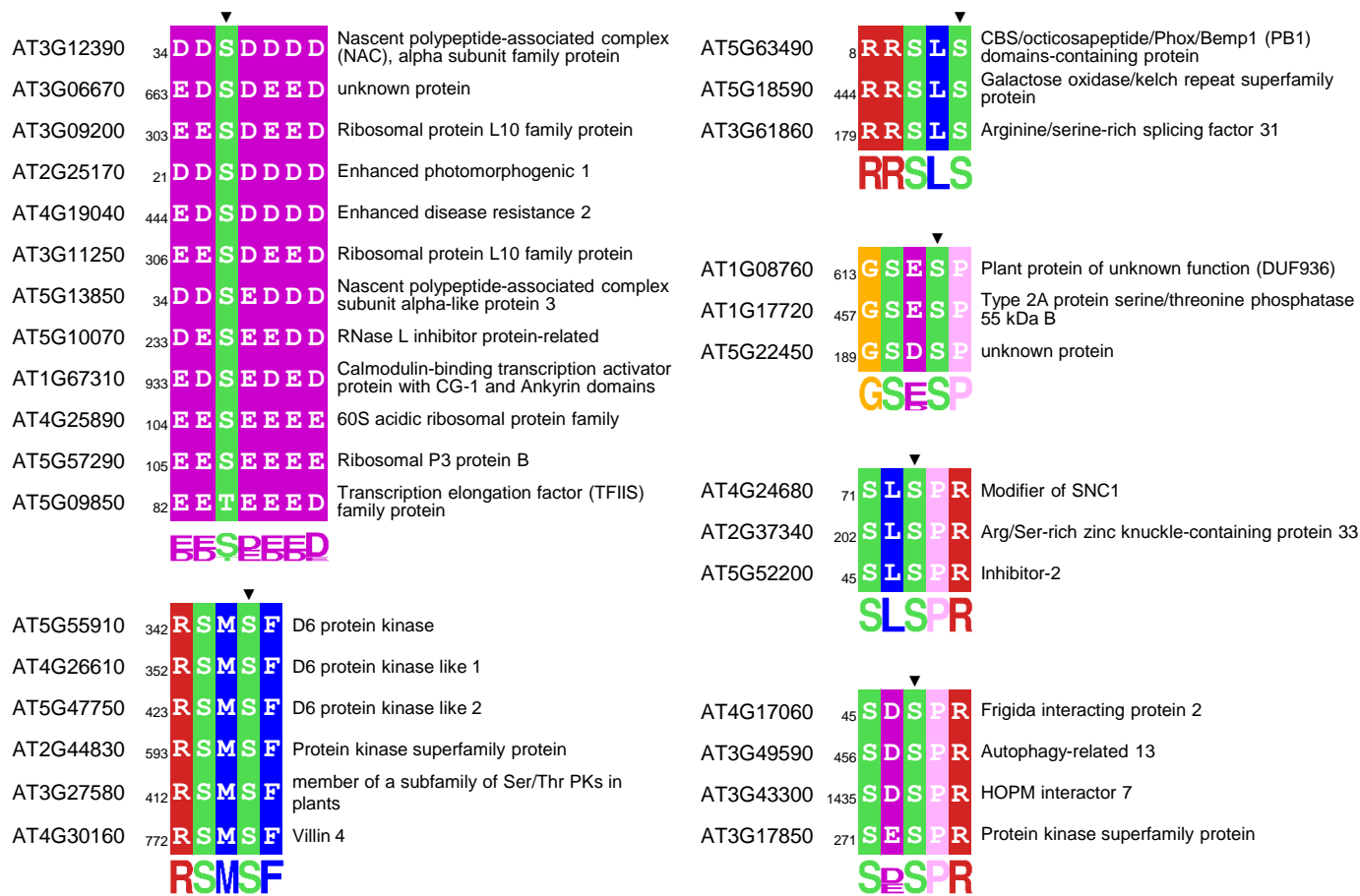

**Fig. S2. Motifs of phosphosites measured by phosphoproteomic analysis of touch-treated and untouched Arabidopsis.**

All of these phosphopeptides used for the construction of conserved phosphosite motifs were discovered from this phosphoproteomics. Alignment was performed using Motif-X followed by MEME, and motif was depicted using LOGO. The Arabidopsis Information Resource accession number and annotation are labeled on the right and the left side of these phosphosites, respectively. The numbers at subscripts of peptide sequences indicate positions of the first amino acid residues in their corresponding proteins. Triangles (▼) denote the phosphosites of phosphoproteins. It is related to Fig. 2 and Dataset S1-S3.

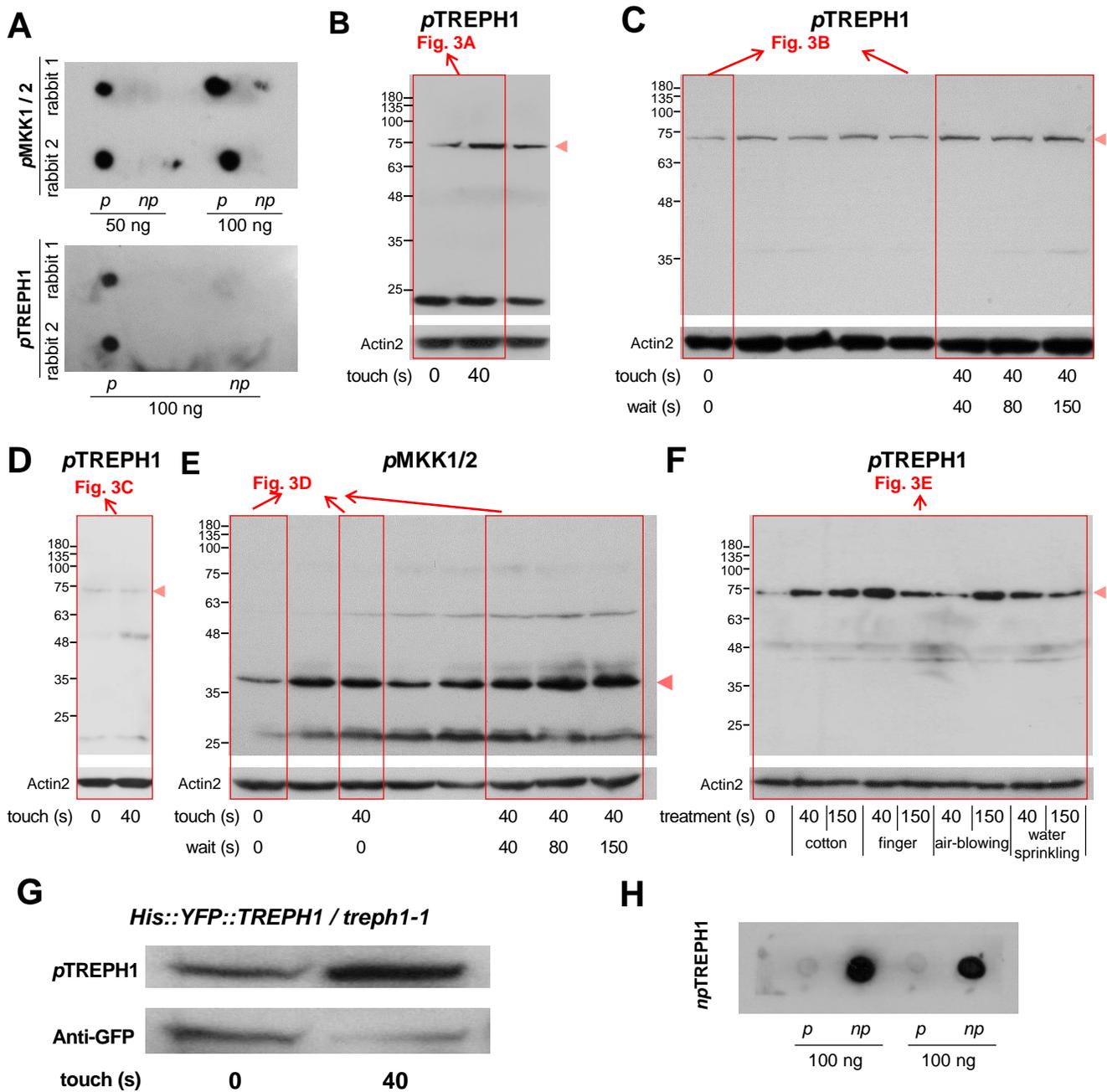

**Fig. S3. Immunoblot analysis of phosphopeptides, TREPH1 and MKK1/2 proteins using phosphosite-specific or protein-specific polyclonal antibodies.**
(*A*) The dot immunoblots of phosphorylated (*p*) and non-phosphorylated (*np*) control peptides using both anti-*p*MKK1/2 and anti-*p*TREPH1 antibodies. (*B-F*) The immunoblots of the whole protein gel showing both the 40-s touch-enhanced phosphorylation at S625 site of TREPH1 and T31 site of MKK2 (and/or T29 of MKK1). The protein bands in the composite Fig. 3A (*B*), 3B (*C*), 3C (*D*), 3D (*E*), 3E (*F*) were excised from these corresponding whole gel lanes (marked in red boxes), respectively. The arrows at the top of gels indicate the original gel lanes, from which those excised protein bands in Fig. 3 were produced. (*G*) The immunoblot analysis of the touch-regulated phosphorylation of transgenic His::YFP::TREPH1 fusion protein. The 3-week-old plants grown in soil were touched 0 s and 40 s, respectively. His::YFP::TREPH1 fusion protein was purified from the total protein from both the control and touched transgenic plants using Ni²⁺-NTA beads. The eluted fusion proteins were used for the immunoblotting analysis. **h,** The dot immunoblots of phosphorylated (*p*) and non-phosphorylated (*np*) control peptides using anti-*np*TREPH1 antibodies. It is related to Fig. 3.

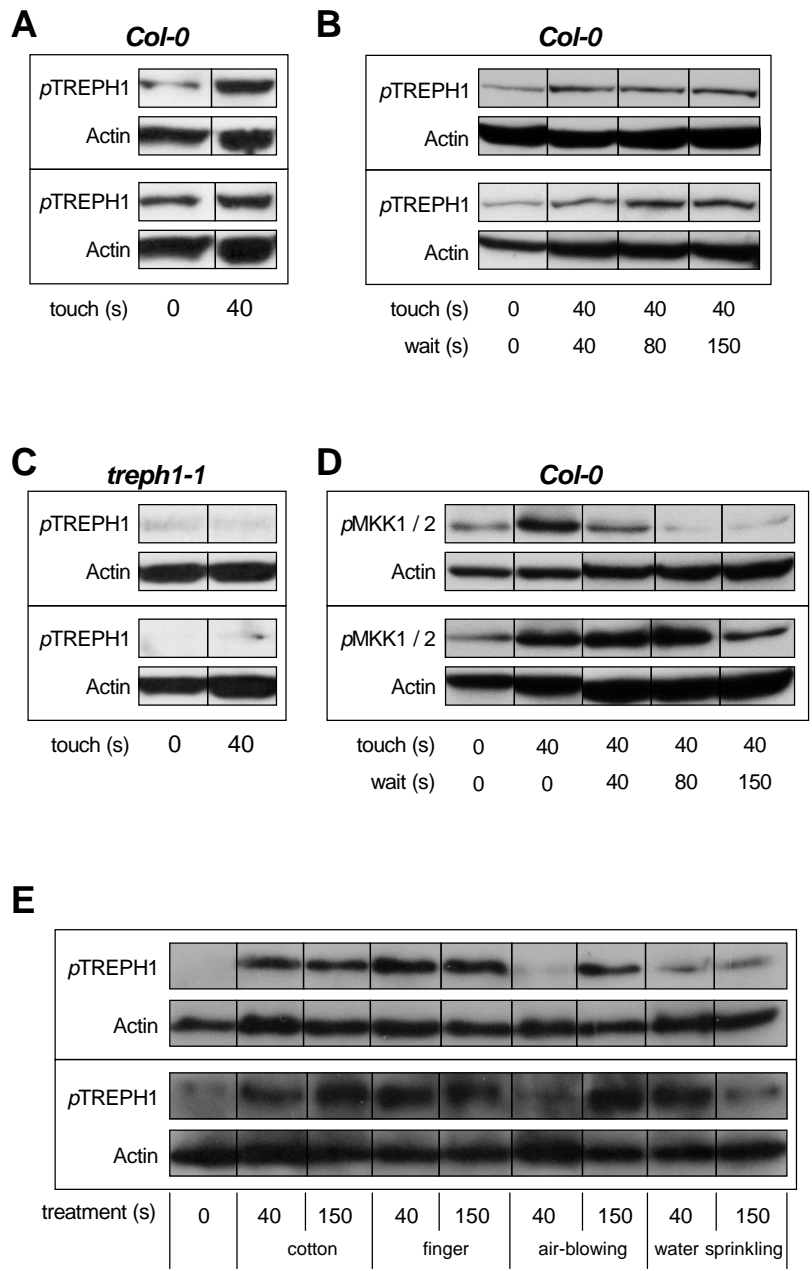

**Fig. S4. Immunoblot analysis of the force-regulated protein phosphorylation.**
(*A-E*) The other two replicates of protein immunoblots on S625 of TREPH1 and T31 site of MKK2 (and/or T29 of MKK1), which were used in the quantitative results of Fig. 3A (*A*), 3B (*B*), 3C (*C*), 3d (*D*) and 3E (*E*), respectively. It is related to Fig. 3.

**A**

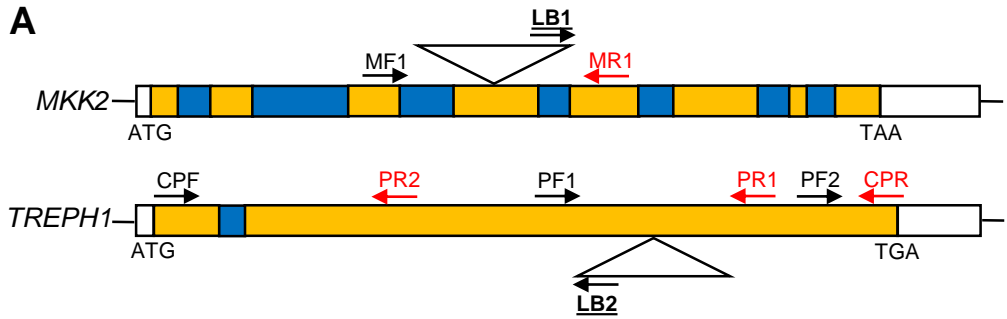

**B** *mkk2*

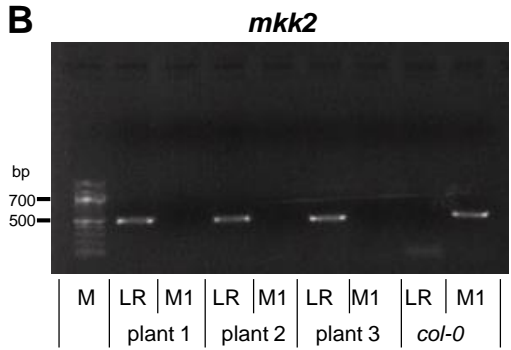

**C** *treph1-1*

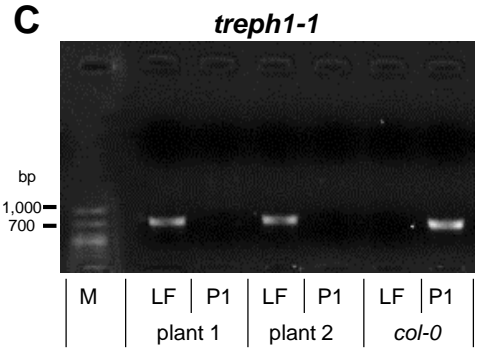

**D**

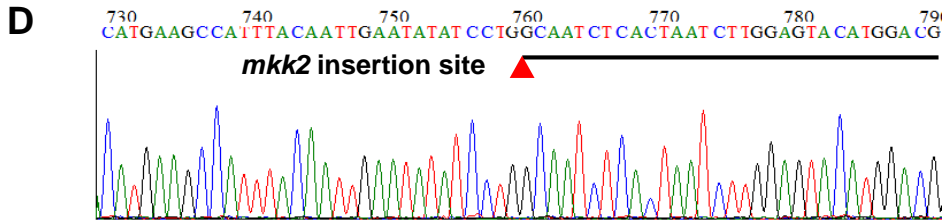

**E**

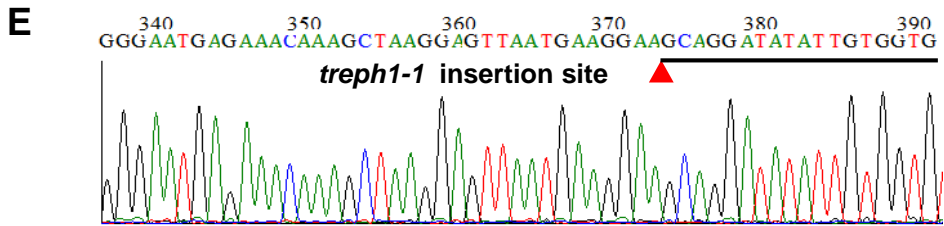

**F**

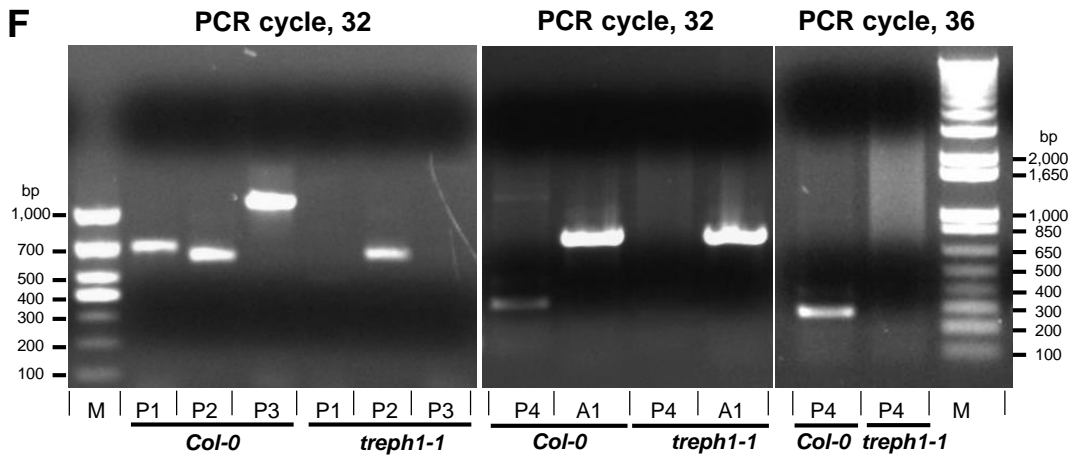

**Fig. S5. Genotyping and transcript measurement of T-DNA insertional mutants.**
(*A*) DNAs schematic representation of *MKK2* and *TREPH1*. Triangles represent T-DNA insertion sites in T-DNA insertional mutants (labeled with plant 1, 2 and 3). Primers used for genotyping and transcripts detection present with arrows. Black and red arrows stand for forward and reverse primers, respectively. Left T-DNA border primers were highlighted with bold and underline. Empty box, 5`- and 3`-UTR. Yellow box, exon region. Blue box, intron region. ATG and TAA/TGA indicate the start and stop codon, respectively. (*B*) Genotyping of *mkk2* T-DNA insertions by PCR. Primer pairs used to amplify each fragment were LB1 + MR1 (LR fragment) and MF1 + MR1 (M1 fragment), respectively. The size of M1 fragment is 554 bp. (*C*) Genotyping of *treph1-1* T-DNA insertions by PCR. Primer pairs used to amplify each fragment were LB2 + PF1 (LF fragment) and PF1 + PR1 (P1 fragment), respectively. The size of P1 fragment is 698 bp. (*D*) T-DNA insertion site verification of *mkk2* by DNA sequencing. The insertion site at 930 bp of *MKK2* gene. (*E*) T-DNA insertion site verification of *treph1-1* by DNA sequencing. The insertion site at 1329 bp of *TREPH1* gene. (*F*) RT-PCR analysis of *TREPH1* transcripts in wild type and *treph1-1* T-DNA insertion line. Primer pairs used to amplify each fragment were PF1 + PR1 (P1 fragment), CPF + PR2 (P2 fragment), PF1 + CPR (P3 fragment) and PF2 + CPR (P4 fragment), respectively. P1 fragment (698 bp) was used for insertion region verification and can only be amplified on wild type. P2 fragment (620 bp) was used for N terminal region verification and can be amplified both on wild type and *treph1-1*. P3 fragment (1097 bp) was used for both insertion and C-terminal region verification which can only be amplified on wild type. P4 fragment (274 bp) was used for C-terminal region verification and can only be amplified on wild type. Primer pair used to amplify fragment for cDNA quality control was ACF + ACR (A1 fragment from *ACTIN2*, AT3G18730). The size of A1 fragment is 721 bp. M, DNA marker. All the primers used are listed in Table S1. It is related to Fig. 3-5.

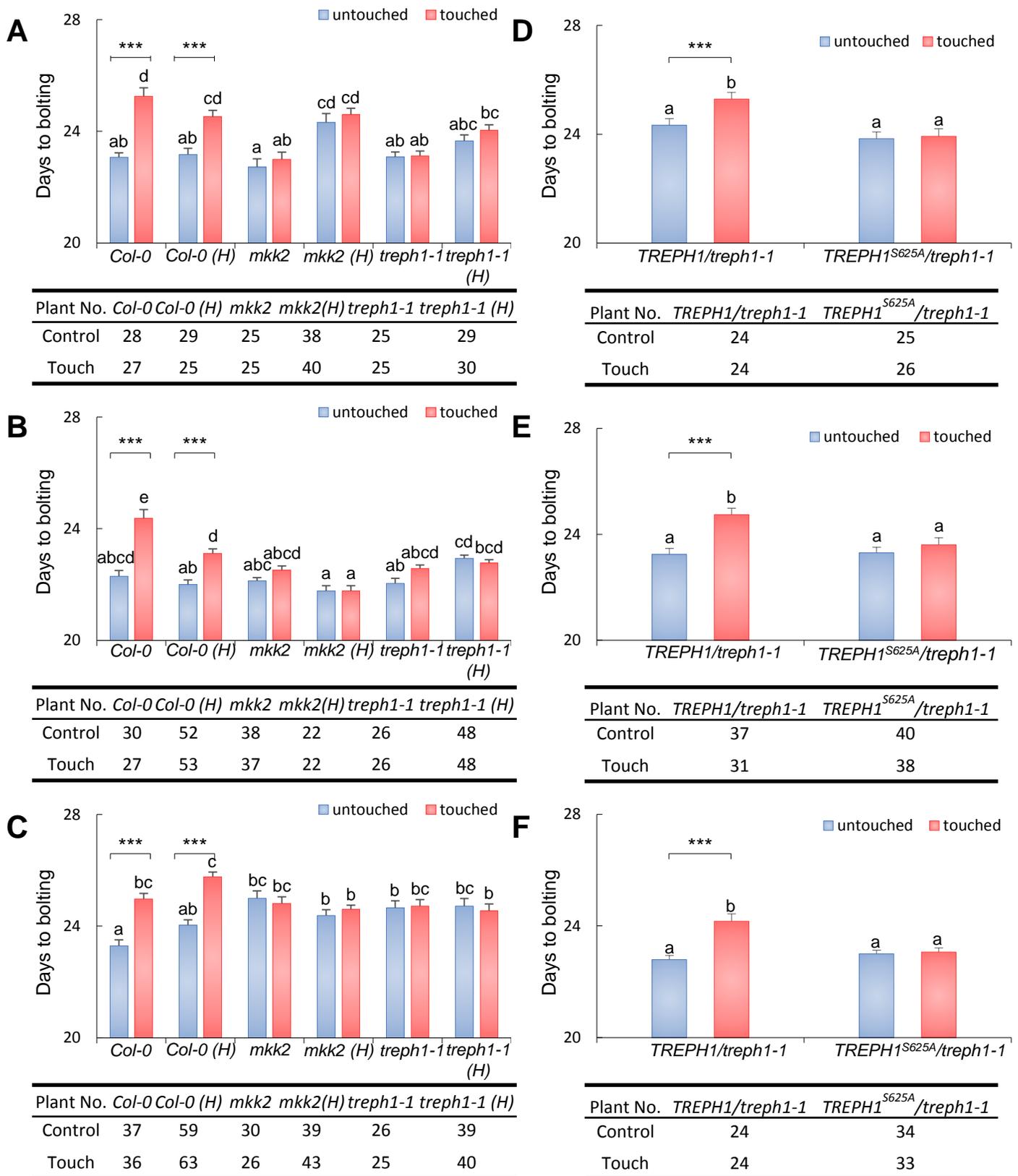

**Fig. S6. Touch responses of wild type and mutants determined by three biological replicates.**
(*A-C*) The bolting times of both the untouched control and the touched wild type and mutant plants from 3 biological replicates. Plants surveyed are shown in the table below. H stands for human hair touch controlled by a fully automated machine. The rest were all touched by cotton swab. (*D-F*) The bolting time of both the untouched control and the touched transgenic plants from 3 biological replicates. Plants surveyed are shown in the table below.
Means ± SE are shown.(n = 3). Statistical test was performed employing student's *t*-test and Tukey's range test. Significance of p < 0.05 and p < 0.01 for pairwise *t*-test are shown as * and **, respectively. The homogeneity of variance in each panel with multiple values was analyzed using Tukey's range test. Different letters represent significant differences at the 5% level. It is related to Fig. 4.

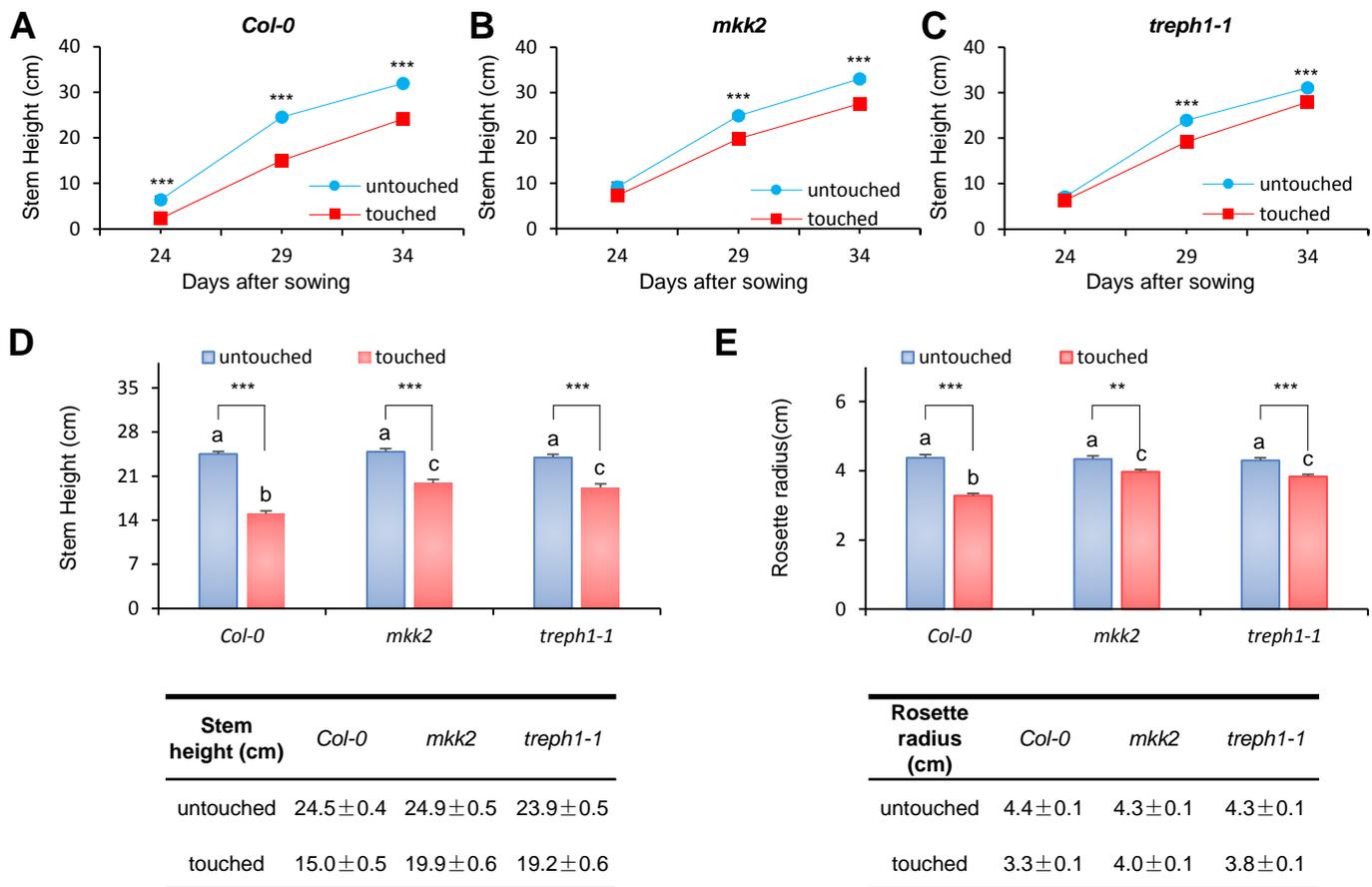

| Stem height (cm) | Col-0 | mkk2 | treph1-1 |
|---|---|---|---|
| untouched | 24.5±0.4 | 24.9±0.5 | 23.9±0.5 |
| touched | 15.0±0.5 | 19.9±0.6 | 19.2±0.6 |

| Rosette radius (cm) | Col-0 | mkk2 | treph1-1 |
|---|---|---|---|
| untouched | 4.4±0.1 | 4.3±0.1 | 4.3±0.1 |
| touched | 3.3±0.1 | 4.0±0.1 | 3.8±0.1 |

**Fig. S7. Morphological measurement of additional Arabidopsis touch-response traits: stem height and rosette size.**
(*A-C*) The stem height of both the control and the touched *col-0* (*A*), *mkk2* (*B*) and *treph1-1* (*C*) by 24, 29 and 34 days after sowing, respectively. (*D*) The stem height results of plants by day 29 after sowing and the table of stem height. (*E*) The average rosette radius by day 29 after sowing (the distance from the rosette center to leaf tip was measured as the rosette radius) and the table of average rosette radius. Touching started on 12-day-old plants. Each plant was touched 3 rounds per day (40 s per round, 1 s per touch). Both the untouched control and the touched *col-0*, *LUC*, *mkk2* and *treph1-1* plants were measured morphologically. Means ± SE are shown (n = 3). All of data were compiled from 2 biological replicates (n ≥ 27, n = number of individuals per replicate). Statistical test was performed employing student's *t*-test and Tukey's range test. Significance of p < 0.01 and p < 0.001 for pairwise *t*-test are shown as ** and ***, respectively. The homogeneity of variance in each panel with multiple (more than two) values was analyzed using Tukey's range test. Different letters represent significant differences at the 5% level. It is related to Fig. 4.

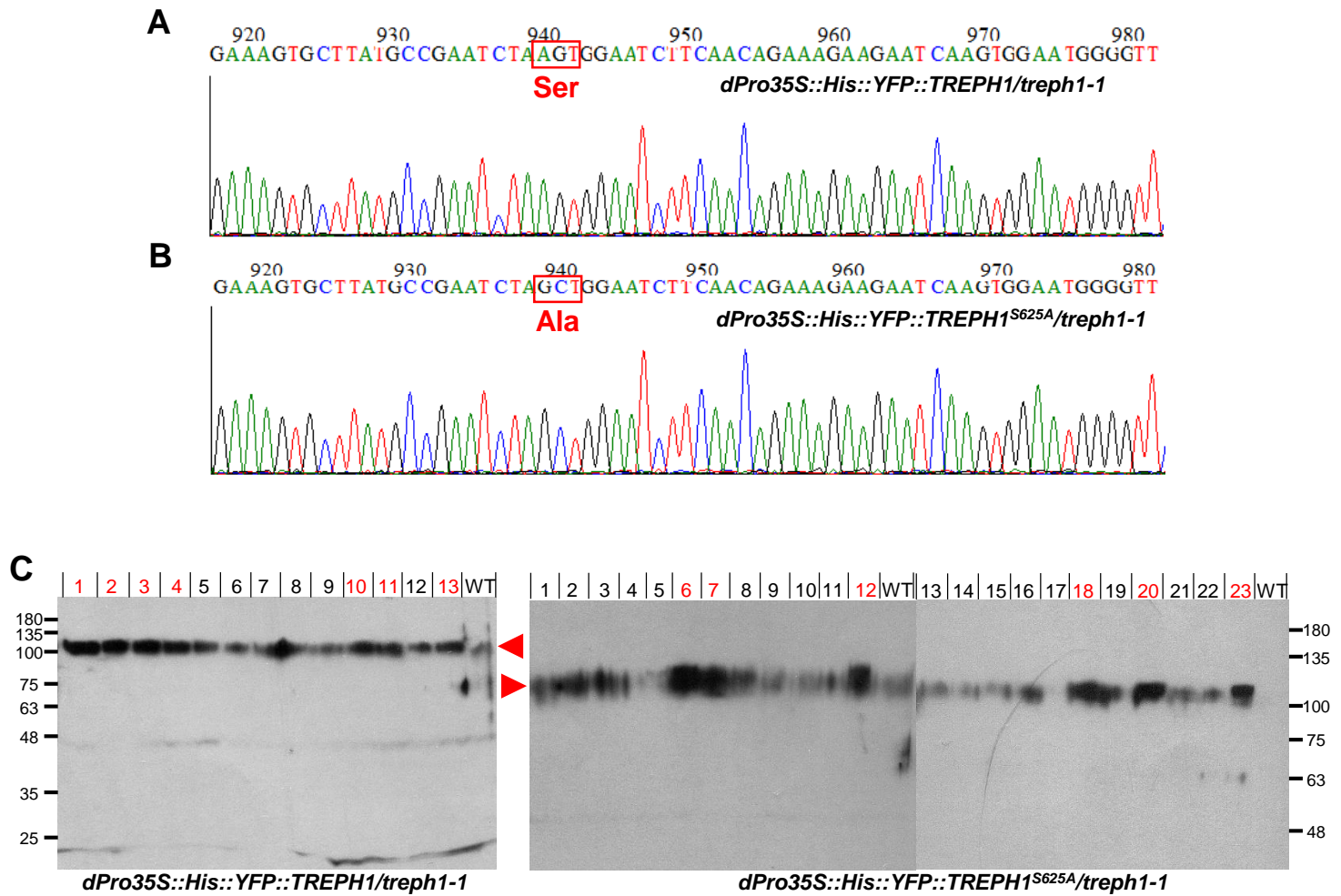

**Fig. S8. Genotyping of Arabidopsis plants overexpressing TREPH1 and TREPH1$^{S625A}$ protein isoforms.**
(*A and B*) DNA sequences of S625 site and S625A mutation site, which encodes a Ser (AGT, *A*) from *TREPH1/treph1-1* and a Ala (GCT, *B*) from *TREPH1$^{S625A}$/treph1-1*, respectively. (*C*) Immunoblot results of T1 transgenic plant selection. In total, we have performed the floral dip on at least 30 individual T$_0$ plants. Seven individual T1 transgenic plants of *dPro35S::His::YFP::TREPH1 / treph1-1* (left, 10% SDS-PAGE) and six individual T1 plants of *dPro35S::His::YFP::TREPH1$^{S625A}$ / treph1-1* right panels, 8% SDS-PAGE) were selected for molecular and functional studies. The selected transgenic lines were marked with red number, whereas red triangles mark the target fusion protein band (His::YFP::TREPH1, ~106KDa). It is related to Fig. 4.

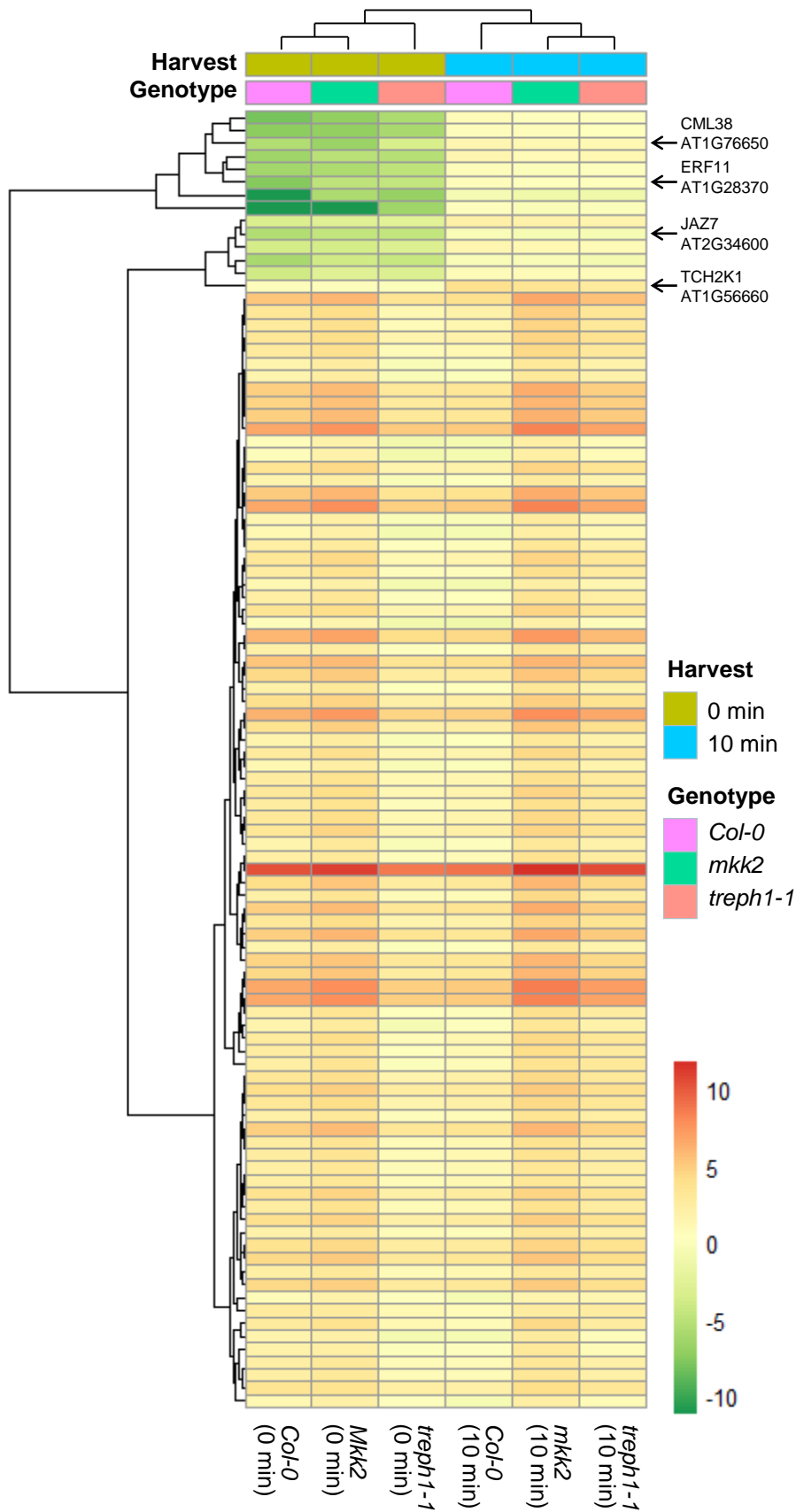

**Fig. S9. Log2-transformed median-centered expression of transcripts showing varied touch-induced alterations between the wild type and *treph1-1*.**

The 3-week-old wild type *Col-0, mkk2* and *treph1-1* mutant plants were touched 40 s, respectively. The tissues of control were harvested immediately after touch (0 min). The tissues of treatment were harvested after a 10-min time lag. The four genes selected for RT-qPCR study were annotated by arrows. Both gene and genotype clustering used Euclidean distance. It is related to Fig. 5.

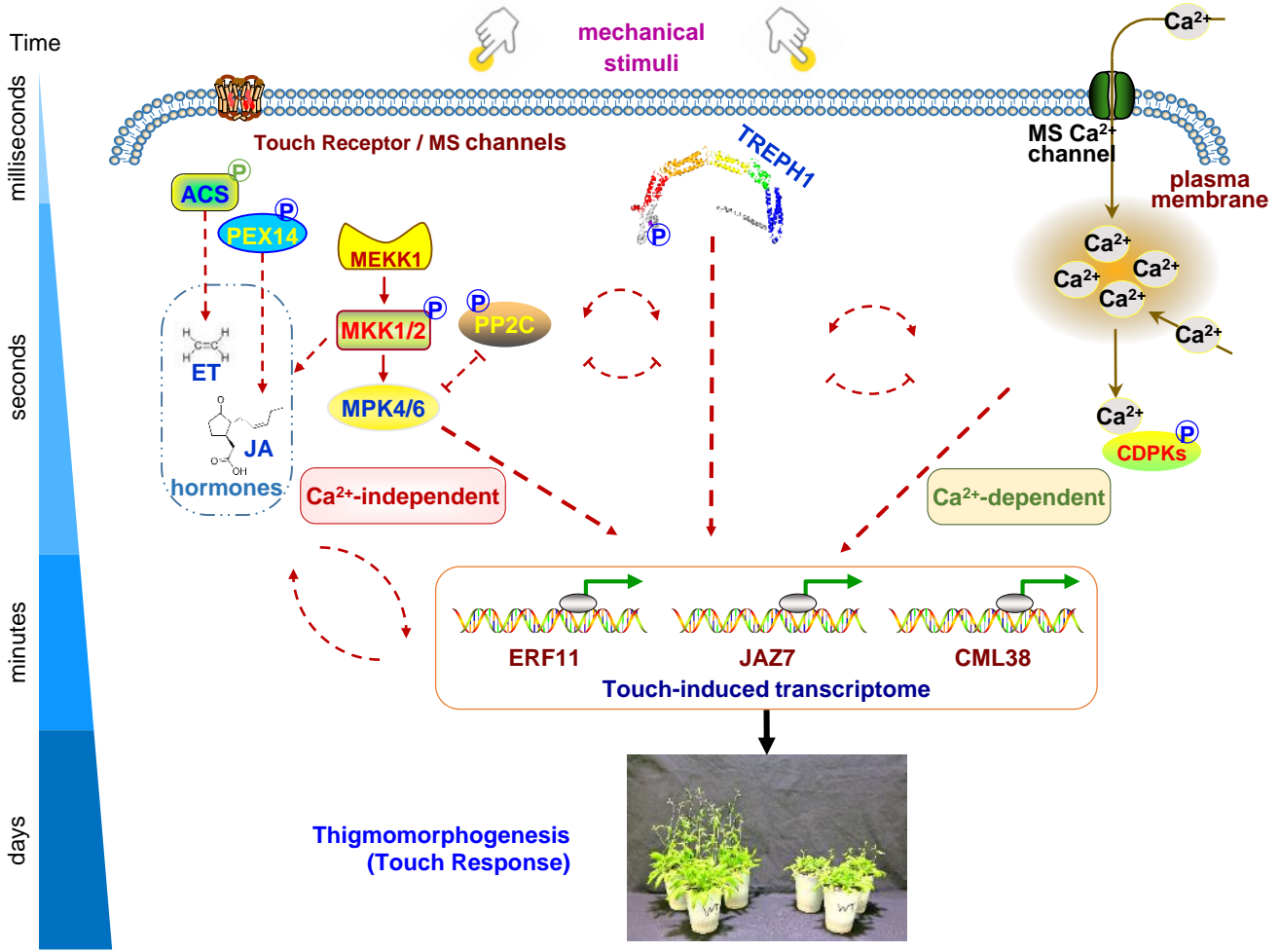

**Fig. S10. Mechanotransduction modeling.**
Mechanistic action model built for touch signaling based on information collected from both quantitative and molecular PTM proteomics and transcriptomics analysis. TREPH1, MKK1/2, and CDPKs transduce mechanical signals to regulate *CML38*, *ERF11*, and *JAZ7* gene expression. ET, ethylene; JA, jasmonic acid. Phosphoproteins identified from the present and previous studies are labeled with blue and green P's in circles, respectively. Solid red lines, direct interaction; dashed red lines, uncharacterized effects; yellow lines, movement of $Ca^{2+}$. Arrow, positive effect; horizontal line at the ends of the dashed line, inhibitory effect. The triangle on the left side indicates the time scale of the force signaling cascades in touch response. Touch receptors or/and mechanosensitive (MS) ion channels like MscS-Like as well as MS $Ca^{2+}$ channels (MCA) are marked on the plasma membrane, respectively. It is related to Fig. 6.

**Table S1. All the primers used in this paper.**

Genotyping

| Accession No./ Description | Primer name | Sequence (5'-3') |
|---|---|---|
| pSKI15, left T-DNA border | LB1 | GACGTGAATGTAGACACGTCG |
| pROK2 (SALK line), left T-DNA border | LB2 | GCGTGGACCGCTTGCTGCAACT |
| AT5G55860 | PF1 | AACTGCAGAAGCAAATGGAGA |
| | PR1 | TTCGCTAGCTCTCACTGCTTC |
| AT4G29810 | MF1 | GTGTTGTTCAGCTGGTTCAAC |
| | MR1 | CGATGGATGATATGCCTATCGT |

**Overexpression transgenic lines**

| Transgenic line | Primer name | Sequence (5'-3') |
|---|---|---|
| *ProCML39::LUC/Col-0* | LUF | CACCAAACTTTGCCGGAAACTATCAC |
| | LUR | ATACCCGGGTTTGAGAAAGAAAAGATTGTATTTG |
| *dPro35S::His::YFP::TREPH1/treph1-1* | CPF | TCAT<u>GGCGCGCC</u>ATGGTTGCTAAGAAGGGACGTAG (AscI underlined) |
| | CPR | GGG<u>GAGCTC</u>AAAAGGGTTTCTCTCCAGG (SacI underlined) |
| *dPro35S::His::YFP::TREPH1<sup>S625A</sup>/treph1-1* | PPF | CCGAATCTA**GC**TGGAATCTTC (mutation sites bolded) |
| | PPR | GTTGAAGATTCCA**GC**TAGATTCG (mutation sites bolded) |

**Transcripts of RT-PCR**

| Accession No./ Description | Primer name | Sequence (5'-3') |
|---|---|---|
| AT5G55860 | PF2 | CAGGAGATGGCGAGAAAGAGAT |
| | PR2 | AGCAACTCAATCTTCTCAGAATG |
| AT3G18730 | ACF | CCAAGCTGTTCTCTCCTTGTAC |
| | ACR | TTAGAAACATTTTCTGTGAACG |

**Transcripts of RT-qPCR**

| Accession No./ Description | Primer name | Sequence (5'-3') |
|---|---|---|
| AT1G76650 | CMF | GCTGCTGTTAGATTGTCTGATACGG |
| | CMR | GCTCCATCTTCTTCTCTTCTTCGTC |
| AT2G34600 | JAF | CAAAGAGATGGAGATGCAAACAA |
| | JAR | CGTCCAACGAGCTATGGAAA |
| AT1G28370 | ERF | CGGTGGTGATGGATGTCGTTAG |
| | ERR | AGTTCTCAGGTGGAGGAGGGAAA |
| AT1G56660 | TCF | AGGGAAAGAAAGGAAAGGGAGAG |
| | TCR | GGCGGCTTCATCATCCATC |
| AT1G76640 | CLF | GGCTGTGTTTGCTTACATGGA |
| | CLR | AGTTTAACCGCAGCCTCGG |
| AT5G55860 | PF3 | CAGGAGATGGCGAGAAAGAGAT |
| | PR3 | TGTAATGTTGTTGCGGTGATGA |
| AT4G29810 | MF2 | AATCGTTGGAAACAAGTACGGAAA |
| | MR2 | TCGGAGGTGCATAAGGGAAC |
| AT5G25760 | UBF | GCTTGGGAGTCCTGCTTGG |
| | UBR | TCCCTGAGTCGCAGTTAAGA |

**Video S1. Cotton swab touch treatment and harvesting of tissues grown on agar for proteomic analysis.**

Touch treatment was applied to 3-week-old adult plants grown on agar. Touch was applied for 40 times (1 time/s) with a cotton swab (touch strength, 84 $\pm$ 80 mg). The aerial rosette leaves were touched randomly. Liquid nitrogen was poured onto the plants within one second of the touch stopping. Tissues were then stored at –80 ° C for proteomics analysis. One set of tissues (one glass jar of touch and one glass jar of control) was poured with liquid nitrogen and harvested simultaneously.

**Video S2. Cotton swab touch treatment of plants grown in soil for biochemical, physiological, morphological and genetic studies.**

Plants were grown in the soil to mimic the environmental condition. There were three plants grown in one cup. Touch treatment was applied on 12-day-old plants and lasted for around 18-25 days until all plants bolted. Touch was applied for 40 times (1 time/s) with a cotton swab and three plants were touched sequentially. The aerial rosette leaves were touched randomly. Each plant was touched 3 rounds/day with 6 hours interval. Plants were counted for bolting when the primary inflorescence height reached to 1cm tall.

**Video S3. Human hair touch treatment of plants grown in soil for physiological, morphological and genetic studies.**

Plants were grown in the soil, which can mimic the environmental conditions. Four plants were grown in one cup and touch treatment was performed on 12-14 days-old plants which lasted for around 18-25 days until all plants bolted. Touch treatment was performed with human hair controlled by a fully automated in-house-built machine (HKUST Laboratory Service). Touch treatment was performed 3 rounds per day with 8 hours interval. In each round, the individual plant was touched 40 times like the case of cotton swab touch. Plants were considered to have bolted when the primary inflorescence height reached to 1 cm tall.